\documentclass[a4paper,11pt]{article}
\usepackage{jheppub} 
\usepackage[T1]{fontenc} 
\usepackage[utf8]{inputenc}
\usepackage{braket}
\usepackage{cancel}
\usepackage{colortbl} 
\usepackage{tikz-feynman} 
\usepackage{amsmath}
\usepackage{yfonts}
\allowdisplaybreaks
\definecolor{Orange}{cmyk}{0,0.61,0.87,0}
\definecolor{JungleGreen}{cmyk}{0.99,0,0.52,0}
\definecolor{OliveGreen}{cmyk}{0.64,0,0.95,0.40}
\definecolor{Brown}{cmyk}{0,0.81,1,0.60}
\definecolor{RoyalBlue}{cmyk}{0.71,0.53,0,0.12}
\definecolor{Gray}{cmyk}{0,0,0,0.40}
\definecolor{LightPink}{cmyk}{0.0,0.25,0,0}
\definecolor{LLightPink}{cmyk}{0.0,0.10,0,0}
\definecolor{LightBlue}{cmyk}{0.25,0,0,0}
\definecolor{LightGray}{cmyk}{0,0,0,0.2}
\definecolor{LightGreen}{cmyk}{0.3,0,0.3,0}
\def\nn{\nonumber}
\def\l{\left}
\def\r{\right}

\def\green#1{\textcolor{OliveGreen}{#1}}

\subheader{
\today \hfill IPMU20-0077, \,TUM-HEP 1269/20
}
\title{
	\boldmath Direct detection of vector dark matter through electromagnetic multipoles}
\author[a,b,c]{Junji Hisano}
\author[d]{Alejandro Ibarra}
\author[e,f]{Ryo Nagai}
\affiliation[a]{Kobayashi-Maskawa Institute for the Origin of Particles and the Universe, Nagoya University, Furo-cho Chikusa-ku, Nagoya, 464-8602 Japan}
\affiliation[b]{Department of Physics, Nagoya University,
Furo-cho Chikusa-ku, Nagoya, 464-8602 Japan}
\affiliation[c]{Kavli IPMU (WPI), UTIAS, University of Tokyo, Kashiwa, 277-8584, Japan}
\affiliation[d]{Physik-Department, Technische Universit\"at M\"unchen, James-Franck-Stra\ss{}e, 85748 Garching, Germany}
\affiliation[e]{Dipartimento di Fisica e Astronomia, Universita' degli Studi di Padova, Via Marzolo 8, 35131 Padova, Italy}
\affiliation[f]{Istituto Nazionale di Fisica Nucleare (INFN), Sezione di Padova, Via Marzolo 8, 35131 Padova, Italy}
\emailAdd{hisano@eken.phys.nagoya-u.ac.jp,
ibarra@tum.de,
nagai@pd.infn.it}
\abstract{Dark matter particles, even if they are electrically neutral, could interact with the Standard Model particles via their electromagnetic multipole moments. In this paper, we focus on the electromagnetic properties of the complex vector dark matter candidate, which can be described by means of seven form factors. We calculate the differential scattering cross-section with nuclei due to the interactions of the dark matter and nuclear multipole moments, and we derive upper limits on the former from the non-observation of dark matter signals in direct detection experiments. We also present a model where the dark matter particle is a gauge boson of a dark $SU(2)$ symmetry, and which contains heavy new fermions, charged both under the dark $SU(2)$ symmetry and under the electromagnetic $U(1)$ symmetry. The new fermions induce at the one loop level electromagnetic multipole moments, which could lead to detectable signals in direct detection experiments.}	
	
\begin{document} 
\maketitle
\flushbottom
\section{Introduction}
\label{sec:Intro}

There is mounting evidence for the existence of dark matter in galaxies, clusters of galaxies and the Universe at large scale (see {\it e.g.} \cite{Bertone:2004pz, Bergstrom:2000pn,Bertone:2010zza}). All the current body of evidence for dark matter arises from its gravitational interactions with ordinary matter. However, it is generically expected from particle physics models that the dark matter particle could have additional interactions with our visible sector apart from gravity. 

A simple possibility is that the dark matter interacts electromagnetically. Obviously, the dark matter must be dark. However, electromagnetic interactions are not precluded, as long as they are sufficiently weak to be compatible with current cosmological and astrophysical observations, as well as with current direct, indirect and collider searches. In fact, in many models the dark matter particle has electromagnetic interactions, provided there exist ``portal'' particles, which interact both with the photon and with the dark matter particle. A renown example is millicharged dark matter~\cite{Holdom:1985ag} (for an overview, see {\it e.g.}~\cite{Davidson:2000hf}). In this case, the ``portal'' particle is a hidden-photon, which interacts with the dark matter particle, as well as with the Standard Model photon (via kinetic mixing). 

Dark matter particles can also interact electromagnetically even if the electric charge is exactly zero, via higher electromagnetic multipole moments~\cite{Bagnasco:1993st,Pospelov:2000bq,Sigurdson:2004zp,Masso:2009mu,Barger:2010gv,Banks:2010eh}. For instance, a Dirac fermion dark matter candidate acquires via loops a magnetic dipole moment, a charge radius, and an anapole moment when it has a Yukawa coupling with an electromagnetically charged scalar, which acts in this case as ``portal'' particle~\cite{Weiner:2012gm,Fukushima:2013efa,Kopp:2014tsa,Ibarra:2015fqa,Primulando:2015lfa,Sandick:2016zut,Herrero-Garcia:2018koq,Hisano:2018bpz}. For Majorana fermion dark matter, only the anapole moment can be generated, since it is the only multipole that violates the charge-conjugation symmetry~\cite{Kayser:1983wm,Radescu:1985wf}. The electromagnetic interactions, despite the loop suppression, can lead to detectable detection rates in a direct detection experiment, and can be crucial for assessing the detection prospects of some dark matter frameworks, notably those where the dark matter interacts only with leptons or with heavy quarks. 

In this paper we focus on the complex vector as dark matter candidate (some explicit models can be found {\it e.g.} in \cite{Servant:2002aq,Cheng:2002ej,Hubisz:2004ft,Birkedal:2006fz,Hambye:2008bq,Hisano:2010yh, Davoudiasl:2013jma, Gross:2015cwa,Karam:2015jta,Choi:2019zeb,Elahi:2019jeo,Abe:2020mph,Nugaev:2020zcv,Elahi:2020urr}). In these frameworks, the dark matter particle interacts at tree level with the Standard Model through the exchange of heavy fermions or heavy scalars (possibly mixing with the Standard Model Higgs). However, spin-1 dark matter particles could also interact electromagnetically with the Standard Model through their multipole moments. In this paper we will investigate in a model independent way the electromagnetic properties of vector dark matter and their implications for direct detection experiments. 

The paper is organized as follows. In Section \ref{sec:TGV} we discuss the general form of the electromagnetic interactions of a vector dark matter particle with a nucleus. In Section \ref{sec:DD} we calculate the implications for direct detection experiments of the vector electromagnetic multipole moments, and we derive upper limits on the various form factors from experiments. In Section \ref{sec:Model} we present a concrete model of vector dark matter with non-zero electromagnetic interactions. Finally, in Section \ref{sec:Summary} we present our conclusions.

\section{Electromagnetic interactions of vector dark matter with nuclei}
\label{sec:TGV}
We consider a massive complex vector field $V^\mu$ with mass $m_V$ as dark matter candidate. The effective interaction Lagrangian of an on-shell vector field $V^\mu$ with the electromagnetic vector field $A^\mu$ was systematically analyzed in ~\cite{Hagiwara:1986vm} (for earlier works, see \cite{Gaemers:1978hg,Gounaris:1996rz}). Keeping terms up to dimension six it reads:
\begin{align}
{\cal L}/e=&
\frac{i g_1^A}{2 m_V^2}
\left[(V_{\mu\nu}^\dagger V^\mu -V^{\dagger\mu}V_{\mu\nu} )\partial_\lambda F^{\lambda \nu}
-V^{\dagger \mu }V^\nu   \Box F_{\mu\nu}
\right]
\nonumber\\
&+\frac{g_4^A}{m_V^2}
V_\mu^\dagger  V_\nu (\partial^\mu \partial_\rho F^{\rho\nu}
+\partial^\nu\partial_\rho F^{\rho\mu})
\nonumber\\
&+\frac{g_5^A}{m_V^2}
 \epsilon^{\mu\nu\rho\sigma}(V^{\dagger}_{\mu} \overleftrightarrow{\partial_\rho} V_\nu)\partial^\lambda F_{\lambda \sigma}
\nonumber\\
&+i \kappa_A V^\dagger_\mu V_\nu F^{\mu\nu}+\frac{i \lambda_A}{m_V^2} V^\dagger_{\lambda\mu} V^\mu_\nu F^{\nu\lambda}
\nonumber\\
&+i \tilde{\kappa}_A V^\dagger_\mu V_\nu {\tilde F}^{\mu\nu}+\frac{i \tilde{\lambda}_A}{m_V^2} V^\dagger_{\lambda\mu} V^\mu_{~~\nu} \tilde{F}^{\nu\lambda},
\label{effective_lag}
\end{align}
where $V_{\mu\nu}=\partial_\mu V_\nu -\partial_\nu V_\mu$, $F_{\mu\nu}=\partial_\mu A_\nu -\partial_\nu A_\mu$, $\tilde{F}_{\mu\nu}=\frac12 \epsilon_{\mu\nu\rho\sigma} F^{\rho\sigma}$, and $(V^\dagger _\mu \overleftrightarrow{\partial_\rho} V_\nu) =
V^\dagger_\mu (\partial_\rho V_\nu)-(\partial_\rho V^\dagger_\mu)  V_\nu$. Further, $e>0$ is the positron charge and $\epsilon^{\mu\nu\rho\sigma}$ is the totally antisymmetric tensor, defined such that  $\epsilon_{0123}=-\epsilon^{0123}=1$.  Here, we impose the $U(1)_{\rm EM}$ gauge invariance and that $V^\mu$ is electrically neutral. For a real vector field $V_{\mu}=V_{\mu}^{\dagger}$, therefore the Lagrangian Eq.~(\ref{effective_lag}) only contains the terms proportional to $g_4^A$ and $g_5^A$. For details, see Appendix \ref{sec:hagiwaraetal}.

\begin{figure}
	\centering
	\includegraphics[width=6cm,clip]{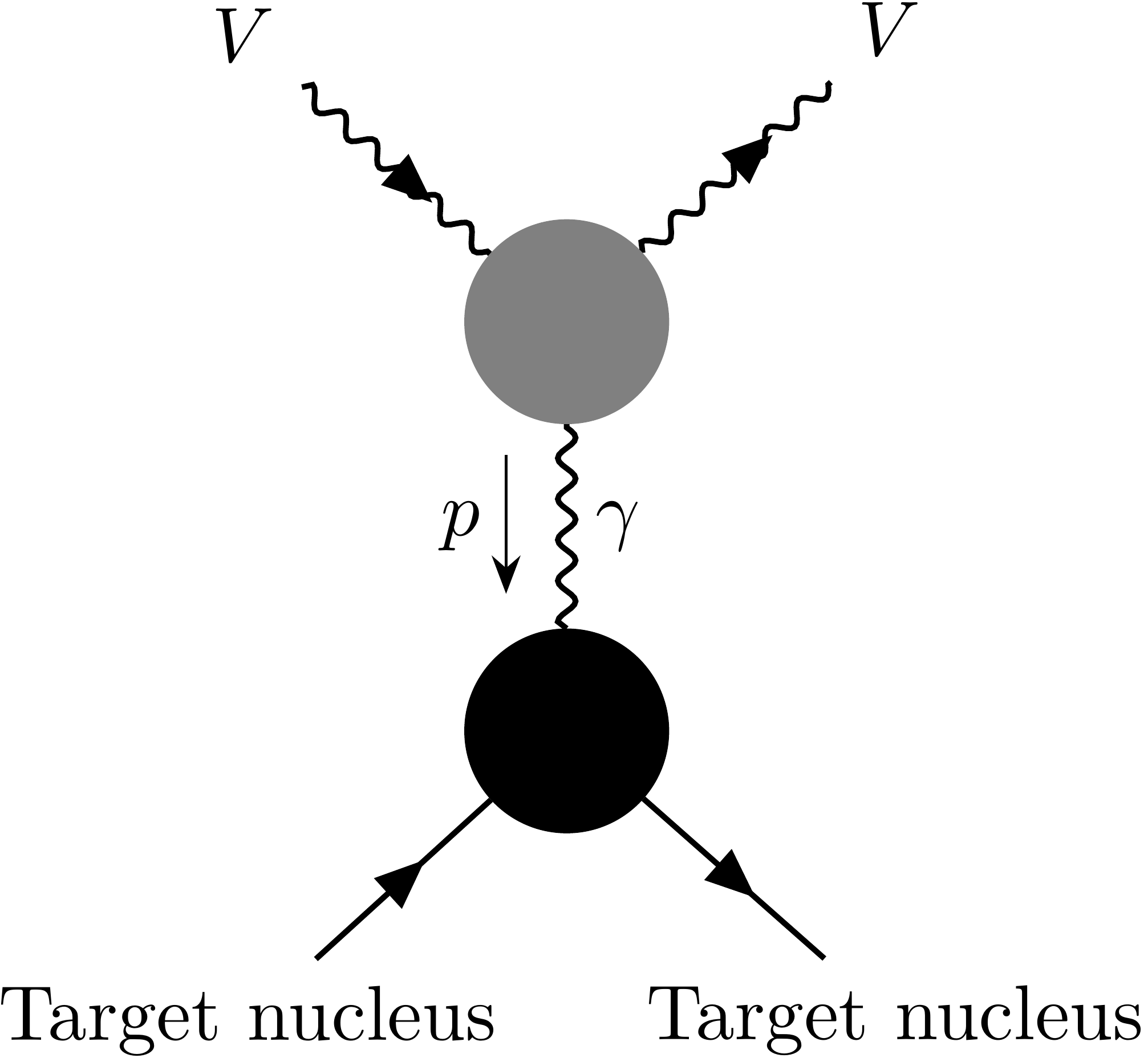}
	\caption{
	Feynman diagram for the scattering of a vector dark matter particle $V$ off a target nucleus due to their electromagnetic multipoles. The gray and black blobs represent respectively the  effective $VV^\dag \gamma$ vertex and the effective electromagnetic interaction vertex of the target nucleus.}
	\label{fig:VNVN}
\end{figure}

The electromagnetic field generated by the vector dark matter particle interacts with the charge and the magnetic moment of the nucleus (see  Figure \ref{fig:VNVN}). We obtain that the differential scattering cross section of the vector dark matter $V$ with a target nucleus with mass $m_T$, atomic number $Z$ and mass number $A$ is:
\begin{align}
\frac{d\sigma}{dE_R}
&=
\frac{Z^2 e^2}{6\pi m_T}F^2_Z(E_R)
\biggl[
\l(\frac{m_T}{E_R}
+\frac{m^2_T-4m_{V} m_T -2m^2_{V}}{4m^2_{V}v^2}
\r)
{(\mu_{V})^2}
+\frac{m_T}{E_Rv^2}
{(d_{V})^2}
\biggr.\nn\\
&
+\frac{3m^2_T}{16v^2}
{(Q_{V})^2}
+\frac{m^2_T}{8}
{(\tilde{Q}_{V})^2}
+\frac{3e^2m^2_T}{4m^4_{V}v^2}
(g^A_1)^2
+\frac{2e^2m^2_T}{m^4_{V}}
(g^A_5)^2
\nn\\
&
\biggl.
+\frac{m^2_T}{4m_{V}v^2}
\mu_{V} Q_{V}
+\frac{3em^2_T}{2m^3_{V}v^2}
\mu_{V} {g}^{A}_1
+\frac{em^2_T}{4m^2_{V}v^2}
Q_{V} {g}^{A}_1
+\frac{m^2_T}{2m_{V}v^2}
d_{V} \tilde{Q}_{V}
\biggr]\nn\\
&+
\frac{e^2}{12\pi m_T}F^2_D(E_R)
\l(\frac{\bar{\mu}_T}{\mu_N}\r)^2
\biggl[\frac{2}{v^2}(\mu_V)^2+(d_V)^2\biggr]\,,
\label{eq:dsigma}
\end{align}
where $E_R$ is the recoil energy (related to the  momentum transfer through $p^2=2 m_T E_R$) and $v$ is the dark matter speed relative to the nucleus. In this expression, we have neglected terms $\mathcal{O}(v^2)$ and $\mathcal{O}(E_R/\textfrak{m}_T)$ (with $\textfrak{m}_T$ the dark matter-nucleus reduced mass) as they only give subdominant contributions to the total rate in direct detection experiments.~\footnote{Such terms arise, for example, from the coupling of the nuclear quadrupole moment (and higher multipoles) to the  term proportional to $g_4^A$ in the Lagrangian Eq.~(\ref{effective_lag}).} 
For convenience, and following \cite{Hagiwara:1986vm}, we have defined
\begin{align}
&\mu_V=\frac{e}{2m_V}(\kappa_A+\lambda_A)\,,\label{eq:defmuV}\\
&Q_V=-\frac{e}{m^2_V}(\kappa_A-\lambda_A)\,,\\
&d_V=\frac{e}{2m_V}(\tilde\kappa_A+\tilde \lambda_A)\,,\\
&\tilde{Q}_V=-\frac{e}{m^2_V}(\tilde\kappa_A-\tilde\lambda_A)\,,
\label{eq:defg5A}
\end{align}
with $\mu_{V}, Q_{V}, d_{V}$ and $\tilde{Q}_{V}$ corresponding to the magnetic dipole, electric quadrupole, electric dipole and magnetic quadrupole for $V$, respectively. $g_1^A$ and $g_5^A$ are related to the electric charge radius and to the anapole moment of $V$, respectively. Some of the form factors can interfere with each other, as a result of their identical transformations under parity, charge conjugation and time reversal. These are summarized in Table \ref{tab:CP}. Clearly, in a theory preserving one of those discrete symmetries, the odd form factors identically vanish. Further, $F_Z$ and $F_D$ are the nuclear charge and magnetic dipole moment form factors~\cite{Helm:1956zz,Lewin:1995rx}: 
\begin{align}
F^2_Z(p^2)
&=
\l(\frac{3j_1(pR)}{pR}\r)^2 e^{-p^2s^2}, \\
F^2_{\rm{D}}(p^2)
&=
\left\{
\begin{array}{ll}
\l[\frac{\sin(pR_D)}{pR_D}\r]^2 & (pR_D<2.55, pR_D>4.5)\\
0.047 & (2.55\leq pR_D\leq4.5)
\end{array}
\right.\,.
\end{align}
where $j_1(x)$ is a spherical Bessel function of the first kind, $R=\sqrt{c^2+\frac{7}{3}\pi^2 a^2-5s^2}$ (with $c=(1.23A^{1/3}-0.60)\,\mbox{fm}$, $a=0.52\,\mbox{fm}$ and $s=0.9\,\mbox{fm}$) and $R_D\simeq 1.0A^{1/3}$ fm. Further,  ${\mu}_N=e/2m_p$ denotes the nuclear magneton,  and $\bar \mu_T$ is the weighted dipole moment for the target nuclei, defined as:
\begin{align}
\bar{\mu}_T
=
\l(
\sum_{i}f_i\mu_{i}^2\frac{S_{i}+1}{S_{i}}
\r)^{1/2}\,,
\end{align} 
where $f_i$, $\mu_{i}$, and $S_{i}$ are the elemental abundance, nuclear magnetic moment, and spin, respectively, of the isotope $i$ \cite{Chang:2010en}. 

We note that the terms proportional to $\mu_V$, $d_V$ are enhanced by a factor $1/E_R$, and the terms proportional to $\mu_V$, $d_V$, $Q_V$, $g_1^A$ by a factor $1/v^2$. The term proportional to $d_V$ is doubly enhanced by $1/(E_R v^2)$. These enhancements have important implications for direct detection experiments, as we will discuss in the next section. 

\begin{table}
 \begin{center}
  \def\arraystretch{1.2}
 \setlength{\tabcolsep}{20pt}
\caption{${\bf{C}}$, ${\bf{P}}$, and ${\bf{CP}}$ properties of the various vector dark matter electromagnetic multipole moments.}
\label{tab:CP}
\vspace{5pt}
\begin{tabular}{c|ccc}
\hline
\hline
 Form factors in Eq.~(\ref{eq:dsigma}) & ${\bf{C}}$ & ${\bf{P}}$ & ${\bf{CP}}$\\
\hline
$\mu_{V},Q_{V}, g^A_1$  & $+$ & $+$ & $+$\\
$d_{V},\tilde{Q}_{V}$  & $+$ & $-$ & $-$\\
$g^A_5$  & $-$ & $-$ & $+$\\
\hline
\hline
\end{tabular}
 \end{center}
\end{table}

\section{Direct detection of vector dark matter through electromagnetic interactions}
\label{sec:DD}

We assume that the dark matter in our Galaxy is in the form of $N$ vectors, $V^i$, $i=1...N$, with mass $m_{V^i}$ and number density in the Solar System $n^i$, such that
\begin{align}
\rho_{\rm loc}
=\sum_i^Nn^im_{V^i}\;.
\label{eq:rhoDM}
\end{align}
We will keep the discussion general and we will not specify how many of these components are real and how many are complex, or whether the complex components are symmetric or asymmetric. In our numerical analysis we will adopt $\rho_{\rm loc}=0.3\,{\rm GeV}\,{\rm cm}^{-3}$.

The differential event rate at a direct detection experiment reads:
\begin{align}
\frac{dR}{dE_R}
=
\frac{1}{m_T}\int d^3v\, v f_{\rm{Lab}}({\vec{v}})
\sum_{i=1}^Nn^i\frac{d\sigma^i}{dE_R}\;,
\label{eq:dRdE}
\end{align}
where  $d\sigma^i/dE_R$ is the dark matter-nucleus differential cross section, discussed in Section \ref{sec:TGV}, and $f_{\rm{Lab}}(\vec{v})$ denotes the DM velocity distribution in the laboratory frame. For the latter, we will adopt a Maxwell-Boltzmann distribution in the galactic frame, truncated at the escape velocity from the Galaxy, $v_{\text{esc}}$:
\begin{equation}
 f_{\rm{Lab}}(\vec{v}) = f (\vec{v} + \vec{v}_\text{E}) ~,
\end{equation}
with $\vec{v}_{\text{E}}$ the velocity of the Earth in the galactic frame and 
\begin{equation}
 f (\vec{v}) = 
\begin{cases}
 \frac{1}{\cal N} e^{-v^2/v_0^2}  & (|\vec{v}| < v_{\text{esc}}) \\ 
 0& (|\vec{v}| > v_{\text{esc}}) 
\end{cases}
~,
\end{equation}
with 
\begin{equation}
 {\cal N} = \pi^{3/2} v_0^3 \biggl[
\text{erf} \biggl(\frac{v_{\text{esc}}}{v_0}\biggr)
- \frac{2 v_{\text{esc}}}{\sqrt{\pi} v_0}  e^{-
\frac{v_{\text{esc}}^2}{v_0^2}} 
\biggr]~.
\end{equation}
Hereafter we take 
$v_{\rm{esc}}=544\,{\rm km}\,{\rm s}^{-1}$,
$v_{0}=220\,{\rm km}\,{\rm s}^{-1}$
and
$v_{\rm E}=232\,{\rm km}\,{\rm s}^{-1}$. 
Finally, we calculate the number of events at a given direct detection experiment integrating $dR/dE_R$ over the recoil energy, taking into account the corresponding detection efficiency.

\begin{figure}
	\centering
	\includegraphics[width=6cm,clip]{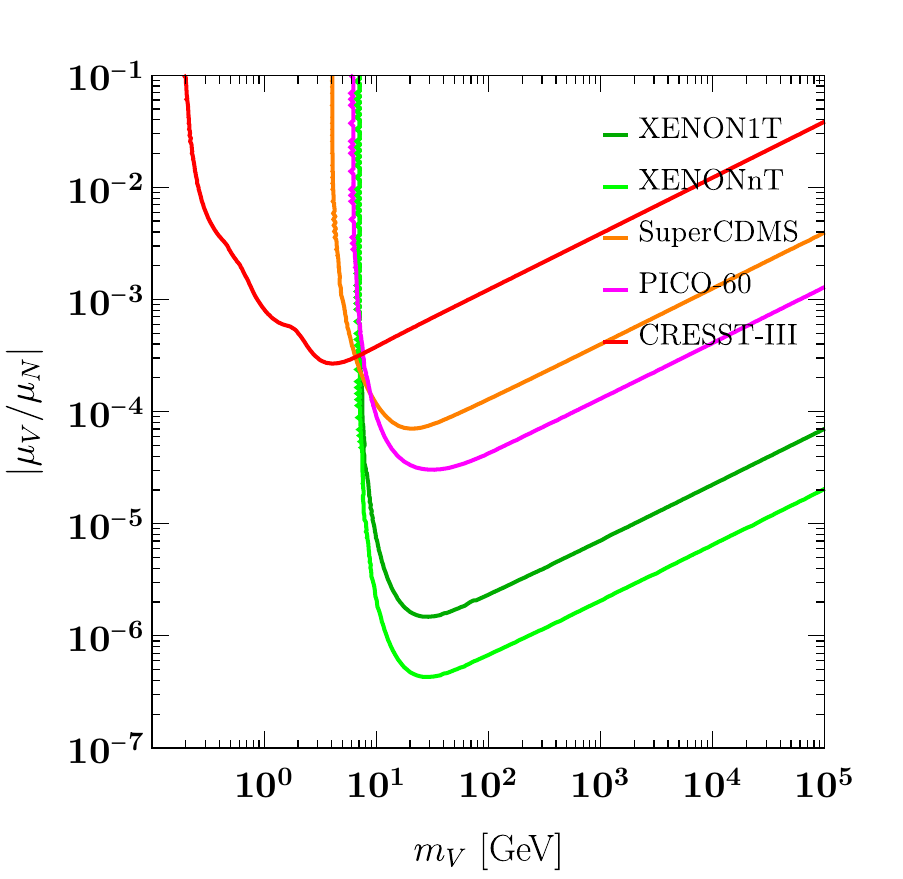}~~~
	\includegraphics[width=6cm,clip]{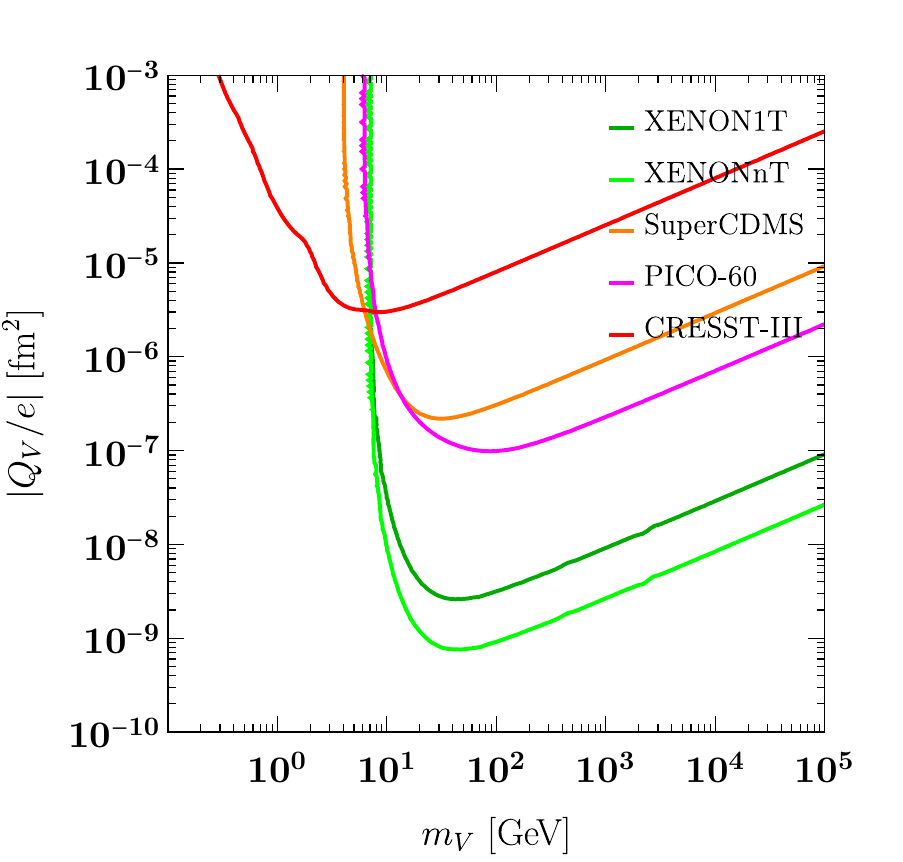}
	\\
	\includegraphics[width=6cm,clip]{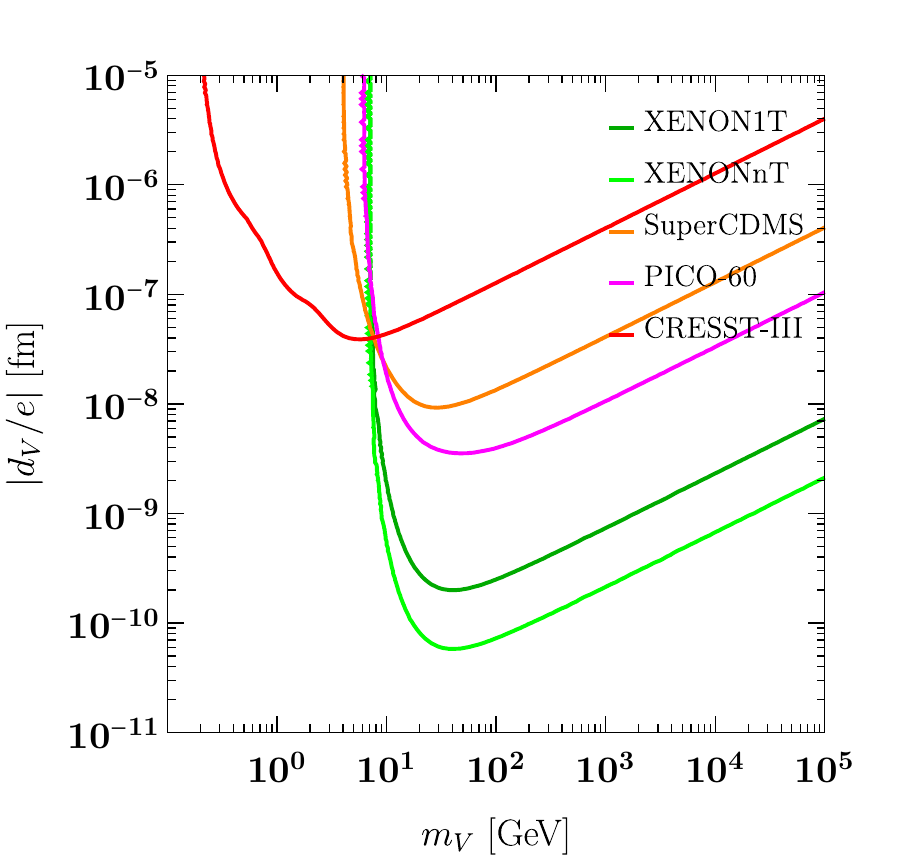}~~~
	\includegraphics[width=6cm,clip]{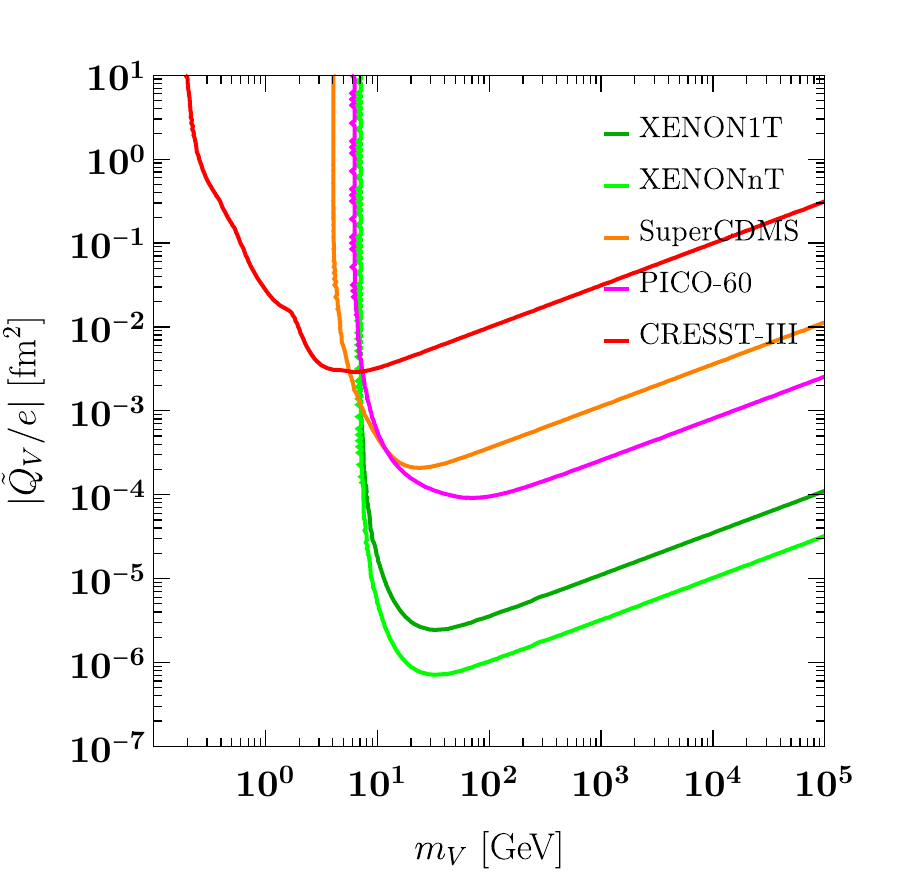}
	\\
	\includegraphics[width=6cm,clip]{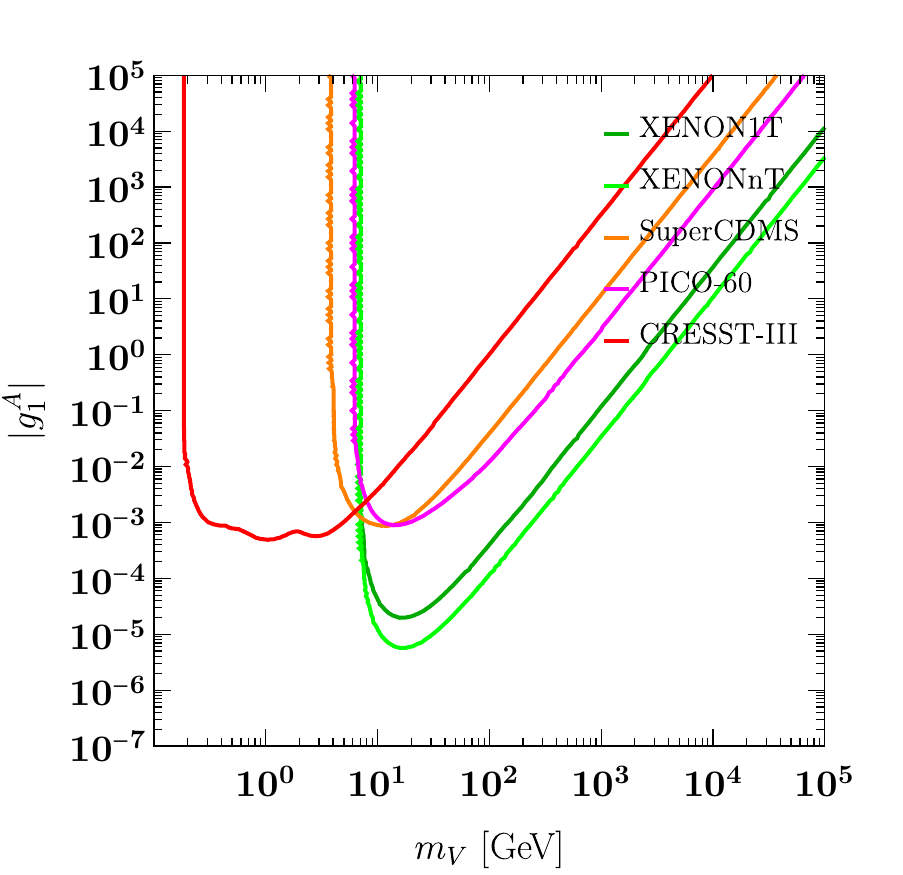}~~~
	\includegraphics[width=6cm,clip]{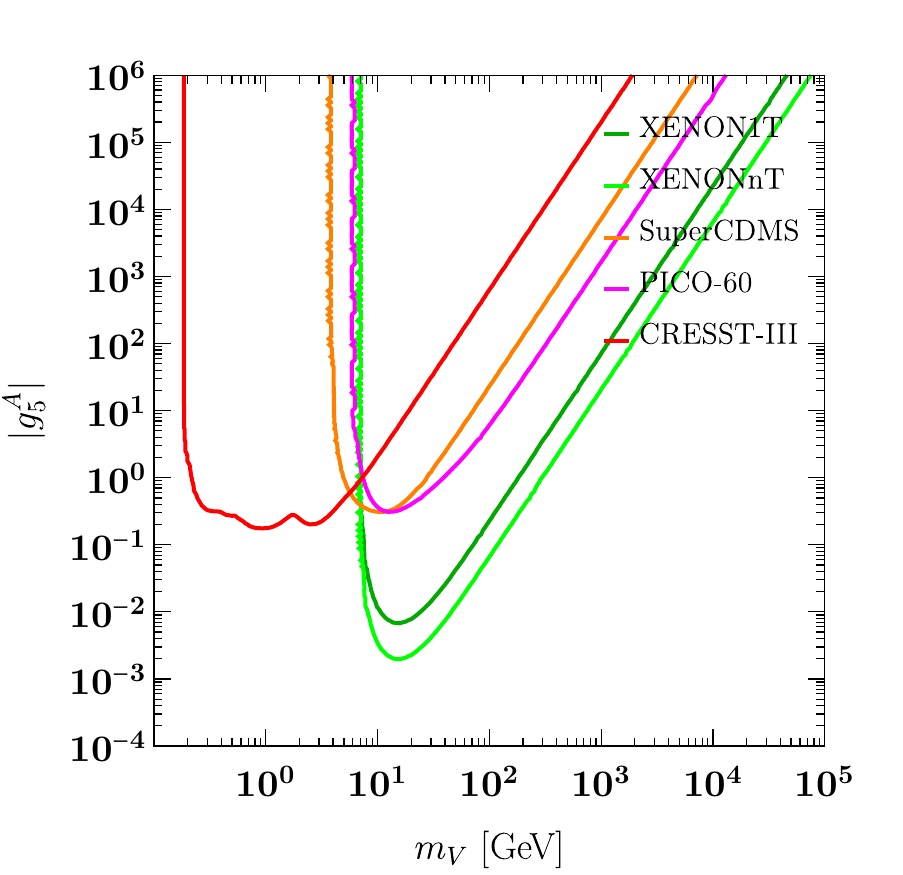}
	\caption{
	Current upper limits in the various vector dark matter electromagnetic multipole moments from the XENON1T, SuperCDMS, PICO-60 and CRESST-III data. We also show the projected sensitivity of the XENONnT experiment.
	}
	\label{fig:DDlimit}
\end{figure}

We show in Figure \ref{fig:DDlimit} the 90\% C.L. upper limits on the various vector dark matter electromagnetic multipole moments from the non-observation of a dark matter signal at the XENON1T \cite{Aprile:2018dbl}, SuperCDMS \cite{Agnese:2014aze}, PICO-60 \cite{Abdelhameed:2019hmk} and CRESST-III \cite{Amole:2019fdf} experiments, alongside with the expected sensitivity of the XENONnT experiment \cite{Aprile:2015uzo}.  Details of the calculation are given in Appendix \ref{sec:DDrate}. For the plots we have assumed that the dark matter is constituted by one complex vector ($V$ and $V^\dagger$), that interacts with the nucleus via one of the form factors in Eq.~(\ref{eq:dsigma}) only. For the real vector, only the form factors $g^A_4$ and $g^A_5$ are non-vanishing. We find that XENON1T sets the most stringent limits for $m_V\gtrsim 10$ GeV, while CRESST-III for $m_V\sim 2-10$ GeV. We also find that the scattering rate is for most experiments  dominated by the dark matter interaction with the nuclear charge, as a result of the enhancement by $Z^2$; the dark matter interaction with the nuclear magnetic dipole moment only plays a role for the PICO experiment.

Let us note that other search strategies could set more stringent limits on the form factors, notably collider search experiments, and could cover the low mass region untested by direct detection experiments. A detailed analysis will be presented elsewhere \cite{inprep}.

\section{A model of vector dark matter with electromagnetic interactions}
\label{sec:Model}

We extend the Standard Model (SM) gauge group with a non-abelian $SU(2)_D$ gauge symmetry and a $U(1)_X$ global symmetry. We assume that the symmetry is spontaneously broken by the vacuum expectation value of a spin-0 field $\Phi_D$, doublet under $SU(2)_D$ and with charge $1/2$ under $U(1)_X$. The vacuum possesses a remnant global $U(1)_D$ symmetry, corresponding to the generator $T_D^3+X$, with $T_D^3$ and $X$ being respectively generators of the $SU(2)_D$ and the $U(1)_X$ symmetries. This remnant symmetry ensures the stability of the lightest among all particles charged under the $U(1)_D$ symmetry.~\footnote{The $U(1)_D$ symmetry is analogous to the electromagnetic $U(1)$ symmetry, which arises after the spontaneous breaking of the $SU(2)_L\times U(1)_Y$ symmetry by the Higgs field, doublet under $SU(2)_L$ and with hypercharge 1/2. The only difference is that we assume the $U(1)_X$ symmetry to be global instead of local, as the $U(1)_Y$ symmetry.} In this case, these are the three massive $W_D$ bosons, which are absolutely stable.

In this simple model the $W_D$ bosons only interact with the Standard Model through the Higgs portal interaction $(H^\dagger H)(\Phi_D^\dagger \Phi_D)$. To couple the $W_D$ bosons to the electromagnetic field, we augment the model with extra fermions, $\Psi_l$ and $\Psi_e$, charged under $U(1)_Y$ and under the dark sector symmetries $SU(2)_D$ and $U(1)_X$. The particle content of the model and the charges under the different symmetry groups are summarized in Table \ref{tab:qnum-fermion}.~\footnote{A similar setup was discussed in Ref.~\cite{Elahi:2019jeo} in the context of multicomponent dark matter scenarios.}

The Lagrangian of the model reads:
\begin{align}
\mathcal{L}
&=
\mathcal{L}_{\rm{SM}}
+
\mathcal{L}_{\rm{kin}}
+
\mathcal{L}_{\rm{mass}}
- V\,,
\end{align}
where
\begin{align}
\mathcal{L}_{\rm{kin}}&=
\bar{\Psi}_li\gamma^\mu
\l(\partial_\mu
+ig_D\sum_{a=1}^3 W^a_{D\mu} T^a
-ig' B_{\mu}
\r)\Psi_l
+\bar{\Psi}_ei\gamma^\mu
\l(\partial_\mu
-ig' B_{\mu}
\r)
\Psi_e\,,\\
-\mathcal{L}_{\rm{mass}}&=
m_{\Psi_l}\bar{\Psi}_lP_L\Psi_l
+
m_{\Psi_e}\bar{\Psi}_eP_L\Psi_e
+\bar{\Psi}_l \Phi_D 
\l(\lambda_L P_L+\lambda_R P_R\r)\Psi_e
+ 
y \bar{\Psi}_l (i\tau^2\Phi^*_D) l_R 
+{\rm h.c.}\,,
\label{eq:yukawa}
\\
V&=
\mu^2_D(\Phi^\dag_D \Phi_D)
+
\frac{\lambda_D}{4}(\Phi^\dag_D \Phi_D)^2
+
\frac{\lambda_{DH}}{4}(\Phi^\dag_D \Phi_D)(H^\dag H)\,
,
\end{align}
where $T^a=\tau^a/2$, with $\tau^a~(a=1,2,3)$ being the $SU(2)_D$ Pauli matrices, and $P_{L,R}=(1\mp \gamma_5)/2$ are the projection operators. $g'$ and $B_\mu$ denote $U(1)_Y$ coupling and gauge boson, respectively. $l_R$ denotes the right-handed SM lepton. Here $g_D$ is real, while $m_{\Psi_l}$, $m_{\Psi_e}$, $\lambda_L$, $\lambda_R$, and $y$ are in general complex quantities. We denote the components of the $SU(2)_D$ doublets as 
\begin{align}
\Psi_l
=
\l(
\begin{array}{ccc}
\Psi_N\\
\Psi_E\\
\end{array}
\r)\,,~~~~~
\Phi_D
=
\l(
\begin{array}{ccc}
\varphi_1+i\varphi_2\\
\varphi_3+i\varphi_4\\
\end{array}
\r)~~~~~
\,.
\end{align}
The Yukawa interaction with coupling $y$ induces the mixing between $\Psi_N$ and the SM lepton after $SU(2)_D$ symmetry breaking. In our numerical analysis, we take $y=0$ for simplicity.

For $\mu_D^2<0$, the $SU(2)_D$ symmetry breaks spontaneously. We work in the gauge where $\langle \varphi_3\rangle=v_D/\sqrt{2}$, $\langle \varphi_{1,2,4}\rangle=0$. Then, the  $W_D$ bosons acquire a common mass $m_{W_D}=g_D v_D/2 $ and the fermion mass terms become:
\begin{align}
-\mathcal{L}_{\rm{mass}}&=
m_{\Psi_l}(\bar{\Psi}_N)_R  ({\Psi}_N)_L
+
\l(
\begin{array}{ccc}
(\bar{\Psi}_E)_R  & ({\bar{\Psi}}_e)_R\\
\end{array}
\r)
\mathcal{M}_E
\l(
\begin{array}{ccc}
({\Psi}_E)_L \\ 
({\Psi}_e)_L
\end{array}
\r)+{\rm h.c.}\,,
\end{align}
where we have defined left- and right-handed fields in the usual manner, $(\Psi)_{L,R}=P_{L,R}\Psi$, and 
\begin{align}
\mathcal{M}_E
=
\l(
\begin{array}{ccc}
m_{\Psi_l} & \frac{\lambda_L}{\sqrt{2}}v_D\\
\frac{\lambda_R}{\sqrt{2}}v_D & m_{\Psi_e}\\
\end{array}
\r)\,.
\label{eq:M_E}
\end{align} 
The mass matrix $\mathcal{M}_E$ can be diagonalized by the field transformation:
\begin{align}
\l(
\begin{array}{ccc}
({\Psi}^1_{E})_L \\ 
({\Psi}^2_{E})_L
\end{array}
\r)
=
V^\dag_L
\l(
\begin{array}{ccc}
({\Psi}_E)_L \\ 
({\Psi}_e)_L
\end{array}
\r)\,,~~~
\l(
\begin{array}{ccc}
({\Psi}^1_{E})_R \\ 
({\Psi}^2_{E})_R
\end{array}
\r)
=
V^\dag_R
\l(
\begin{array}{ccc}
({\Psi}_E)_R \\ 
({\Psi}_e)_R
\end{array}
\r)\,,
\end{align}
where $V_{L,R}$ are unitary matrices satisfying 
\begin{align}
V^\dag_R \mathcal{M}_E V_L
=
\mbox{diag}(m_{E^1},m_{E^2})\,,
\end{align}
with $m_{E^{1,2}}$ being positive and real, and ordered such that $m_{E^1}\leq m_{E^2}$.
The Dirac mass term for $\Psi_N$ is taken to be real and non-negative via an appropriate phase rotation. 
Finally, one finds the following interaction Lagrangian of the mass eigenstate fields with the photon and the $SU(2)_D$ gauge vectors:
	\begin{align}
	\mathcal{L}_{\rm{int}}
	=
	&-\frac{g_D}{\sqrt{2}}
	\biggl(\bar{\Psi}^i_{E}
	\l[(V_L)_{1i} P_L +(V_R)_{1i} P_R \r]\gamma^\mu
	\Psi_N W^-_{D\mu}+h.c.\biggr)
	-e 
	\Psi_N\gamma^\mu \Psi_N A_\mu
	-e 
	\bar{\Psi^i}_{E}\gamma^\mu{\Psi^i}_{E}A_\mu\,.
	\label{eq:eff_Lagrangian_model}
	\end{align}
Generalizations to a larger number of fields are straightforward.

The quantum numbers for the physical particles in this setup are summarized in table \ref{tab:qnum-fermion}.
The states $W_D^\pm=(W_D^1\mp i W_D^2)/\sqrt{2}$, $\Psi^1_E$ and $\Psi^2_E$ transform non-trivially under the remnant $U(1)_D$ and the lightest among them will be absolutely stable. In this work we will assume $m_{E^1}, m_{E^2}>m_{W_D}$, such that $W_D^\pm$ are dark matter candidates ($\Psi_{E_1}$ decays into $W_D$ and SM particles via the Yukawa coupling Eq.~(\ref{eq:yukawa}), with rate proportional to $y^2$). Notice that $W_D^0$ does not carry a $U(1)_D$ charge and can decay. Concretely, the Lagrangian Eq.~(\ref{eq:eff_Lagrangian_model}) induces a kinetic mixing term between $W_D^0$ and $A^\mu$ of the form 
\begin{align}
\mathcal{L}
\supset
{\green{-}}
\frac{\epsilon}{2}W^0_{D\mu\nu} F^{\mu\nu}\,,
\end{align}
 with  $\epsilon$ given by:
\begin{align}
\epsilon
=
\frac{eg_D}{12\pi^2}
\biggl[\log \frac{m^2_{N}}{m^2_{E^2}}
-\frac{1}{2}\Big(
|(V_R)_{11}|^2+|(V_L)_{11}|^2
\Big)\log \frac{m^2_{E^1}}{m^2_{E^2}}
\biggr]
\,,
\label{eq:one-loop-mixing}
\end{align}
Redefining the vector fields in the usual manner to bring the kinetic terms into their canonical form, one finds the coupling in the Lagrangian $\sim e \epsilon J^\mu_{\rm em} W_D^0$, that induces the decay of $W_D^0$ into Standard Model fermions.

\begin{table}
	\begin{center}
		\caption{Particle content and charge assignments for the model present in Section \ref{sec:Model}.}
		\label{tab:qnum-fermion}
		\vspace{5pt}
		\begin{tabular}{c|c|cc|ccc}
			\hline
			\hline
			Gauge eigenstates & Spin &  $SU(2)_D$ &  $U(1)_X$ & $SU(3)_C$ & $SU(2)_L$ & $U(1)_Y$  \\
			\hline
			$W_D$ & 1    & {\bf{3}} & $0 $ & {\bf{1}} & {\bf{1}} & 0 \\
			$\Phi_D$ & 0    & {\bf{2}} & $1/2 $ & {\bf{1}} & {\bf{1}} & 0 \\
			$\Psi_l  $ & 1/2 & {\bf{2}} & $-1/2$ & {\bf{1}} & {\bf{1}} & -1 \\
			$\Psi_e  $ & 1/2 & {\bf{1}} & $-1$ & {\bf{1}} & {\bf{1}} & -1 \\
			\hline
			\hline
			Mass eigenstates & Spin &  $U(1)_D$ & $SU(3)_C$ & & $U(1)_{\rm{EM}}$ &  \\
			\hline
			$W^\pm_D=\frac{1}{\sqrt{2}}(W^1_D \pm iW^2_D)$ &1 & $\pm 1$ & {\bf{1}} & & 0 &  \\
			$W^0_D=W^3_D$ & 1& $0$ & {\bf{1}} & & 0  \\
			$h_D$ &0 & $0$ & {\bf{1}} & & 0  & \\
			$\Psi_N$ &1/2 & $0$ & {\bf{1}} & & $-1$  & \\
			$\Psi^1_{E}$ &1/2 & $-1$ & {\bf{1}} & & $-1$  &  \\
			$\Psi^2_{E}$ &1/2 & $-1$ & {\bf{1}} & & $-1$  &  \\
			\hline
			\hline
		\end{tabular}
	\end{center}
\end{table}

Due to the assignments of gauge charges of the fields $\Psi_l$ and $\Psi_e$, one generically expects these particles to be in thermal equilibrium with the plasma of Standard Model particles. Accordingly, the complex vector dark matter candidates $W_D^\pm$ are also expected to be in thermal equilibrium with the SM. Therefore, for appropriate model parameters $W_D^\pm$ could account for the whole dark matter of the Universe via the mechanism of thermal freeze-out.

The dark matter candidates $W_D^\pm$ in our galaxy interact with the Standard Model particles at tree level via the Higgs portal, or at the one loop-level via the electroweak interactions of the fermions $\Psi_l$ and $\Psi_e$. Direct detection of vector dark matter through the Higgs portal interactions was discussed {\it e.g.} in \cite{Hambye:2008bq}. Here we assume that the Higgs portal interactions are negligibly small, and we focus on the implications for direct detection experiments of the electroweak interactions induced at the quantum level by the fermions  $\Psi_E^{1,2}$. In what follows we will consider only the electromagnetic interactions, discussed in section \ref{sec:DD}, since dark matter interactions with nuclei induced by weak multipoles are expected to be subdominant. 

\begin{figure}
	\centering
	\includegraphics[width=.25\textwidth,clip]{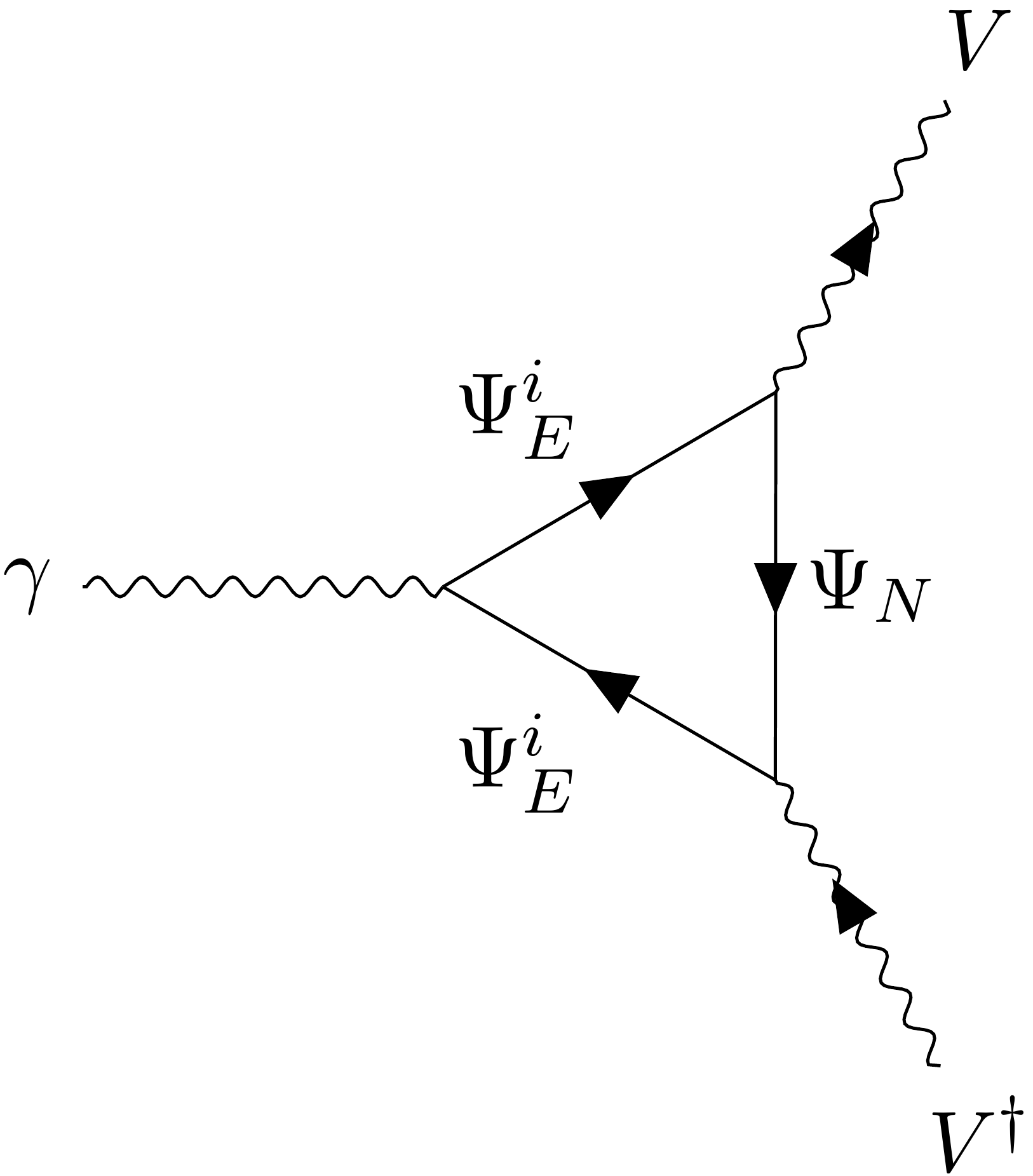}
	\includegraphics[width=.25\textwidth,clip]{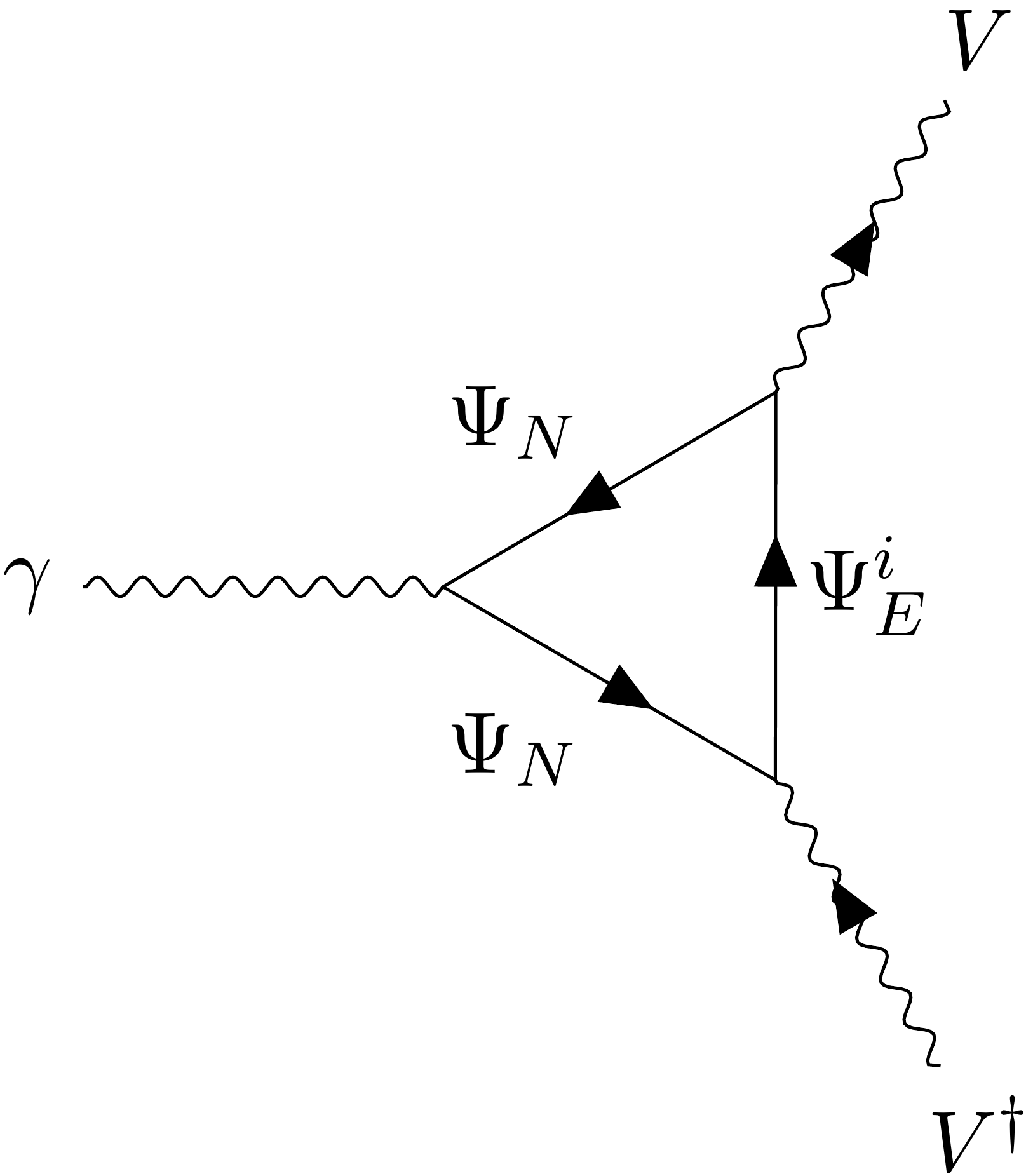}
	\raisebox{.95cm}{\includegraphics[width=.3\textwidth,clip]{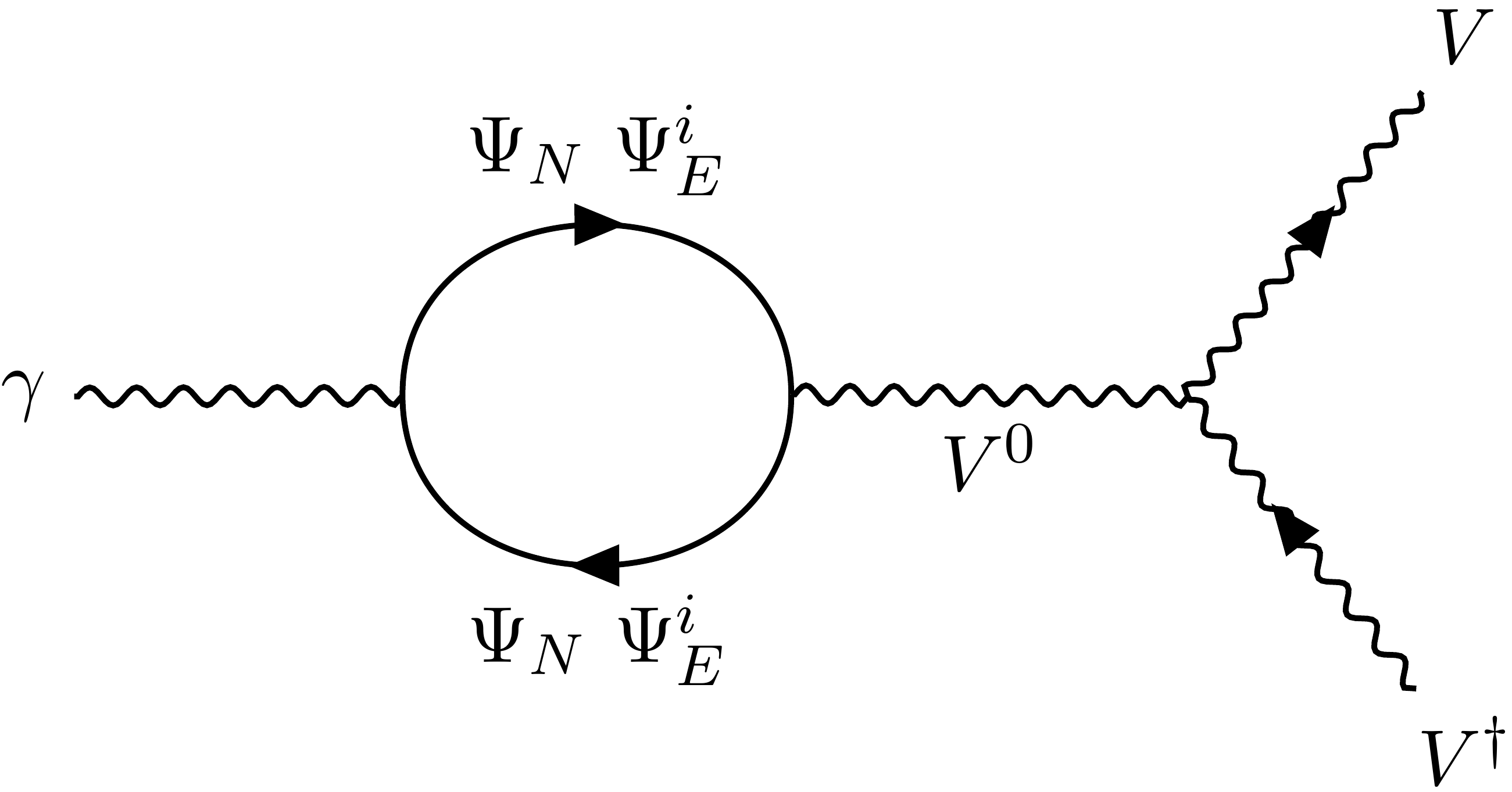}}	
	\caption{One-loop diagrams generating the effective $V^{\dag} V \gamma$ vertex in the model described in Section \ref{sec:Model}.
	}
	\label{fig:loopdiagram}
\end{figure}

The electromagnetic multipole moments can be readily computed from the diagrams in Fig.~\ref{fig:loopdiagram} and read:
\begin{align}
\mu_{V}
&=
-\frac{e{g^2_D}}{64\pi^2m_{W_D^{\pm}}}\sum_{i=1}^2 \Big(\frac{1}{r^2_N}-\frac{1}{r^2_{E^i}}\Big)
\biggl[\biggl(|(V_L)_{1i}|^2+|(V_R)_{1i}|^2\biggr)
\mathcal{G}^{(1)}_{\mu}(r_N,r_{E^i})\nn\\
&\hspace{4cm}
+
2
{\rm Re}
\biggl((V_L)_{1i}^*(V_R)_{1i}\biggr)
\mathcal{G}^{(2)}_{\mu}(r_N,r_{E^i})
\biggr]
\,,
\label{eq:modelmuV}
\\
Q_{V}
&=
-\frac{eg^2_D}{{64\pi^2}m^2_{W_D^{\pm}}}\sum_{i=1}^2 \Big(\frac{1}{r^2_N}-\frac{1}{r^2_{E^i}}\Big)
\biggl[\biggl(|(V_L)_{1i}|^2+|(V_R)_{1i}|^2\biggr)
\mathcal{G}^{(1)}_{Q}(r_N,r_{E^i})\nn\\
&\hspace{4cm}
+
2
{\rm Re}
\biggl((V_L)_{1i}^*(V_R)_{1i}\biggr)
\mathcal{G}^{(2)}_{Q}(r_N,r_{E^i})
\biggr]
\,,\\
g^A_1
&=
-\frac{g^2_D}{{64\pi^2}}
\sum_{i=1}^2 \Big(\frac{1}{r^2_N}-\frac{1}{r^2_{E^i}}\Big)
\biggl[\biggl(|(V_L)_{1i}|^2+|(V_R)_{1i}|^2\biggr)
\mathcal{G}^{(1)}_{1}(r_N,r_{E^i})\nn\\
&\hspace{4cm}
+
2{\rm Re}
\biggl((V_L)_{1i}^*(V_R)_{1i}\biggr)
\mathcal{G}^{(2)}_{1}(r_N,r_{E^i})
\biggr] \nn\\
&\hspace{4cm}
-2 \frac{g_D}{e} \epsilon
\,,
\label{eq:modelg1A}\\
d_V
&=
\frac{eg^2_D}{64\pi^2m_{W_D^{\pm}}}
\,
\sum_{i=1}^2
{\rm Im}\biggl((V_L)_{1i}^*(V_R)_{1i}\biggr)
\mathcal{G}_{d}(r_N,r_{E^i})\,,\\
\tilde{Q}_V
&=
\frac{eg^2_D}{64\pi^2m^2_{W_D^{\pm}}}
\,
\sum_{i=1}^2
{\rm Im}\biggl((V_L)_{1i}^*(V_R)_{1i}\biggr)
\mathcal{G}_{\tilde{Q}}(r_N,r_{E^i})\,,\\
g^A_5
&=
\frac{g^2_D}{128\pi^2}
\,
\sum_{i=1}^2
\biggl(|(V_L)_{1i}|^2-|(V_R)_{1i}|^2\biggr)
\mathcal{G}_{5}(r_N,r_{E^i})\,,
\label{eq:modelg5A}
\end{align}
where $r_N=m_N/m_{W_D^\pm}$ and $r_{E^{i}}=m_{E^i}/m_{W_D^\pm}$. The expressions for the $\mathcal{G}$-functions can be found in Appendix \ref{app:form}.

Some remarks are in order: {\it i)} The vertex diagram induces all form factors, while the wave function renormalization diagram only induces the form factor $f_1(p^2)= g_V \epsilon p^2/(p^2-m^2_{W^\pm})$, proportional to the kinetic mixing parameter $\epsilon$ given in Eq.~(\ref{eq:one-loop-mixing}). This form factor arises from the triple gauge interaction, {\it e.g.} the term $\partial_\mu W_{D\nu}^{+} W_D^{-\mu} W_{D}^{0\nu}$, and from the canonical normalization of the fields $W_D^0$ and $A$, leading to a coupling proportional to $g_1^A$ in Eq.~(\ref{effective_lag}).
{\it ii}) if ${\bf{P}}$ is conserved (such that $V_L=V_R$) then $d_V, \tilde Q_V, g_5^A=0$, and if ${\bf{CP}}$ is conserved (such that all couplings are real) then $d_V,\tilde{Q}_V=0$. This is consistent with the ${\bf{C}}$, ${\bf{P}}$ and ${\bf{CP}}$ transformation properties of the different form factors listed in Table \ref{tab:CP}. {\it iii)} The ${\bf{C}}$ and ${\bf{P}}$ conserving electromagnetic multipoles ($\mu_V$, $d_V$, and $g^A_1$) vanish when the masses of the particles in the loop are degenerate, $m_{N}=m_{E^i}$. This fact was emphasized by Ref.~\cite{Lahanas:1994dv}. {\it iv)} For fixed $r_N, r_{E^i}$, the dipole moments ($\mu_V$, $d_V$) scale as $m^{-1}_{W_D^\pm}$, the quadrupole moments ($Q_V$, $\tilde{Q}_V$) as $ m^{-2}_{W_D^\pm}$ and $g^A_1$ and $g^A_5$ are independent of $m_{W_D^\pm}$.

\begin{figure}
	\centering
	\includegraphics[width=6cm,clip]{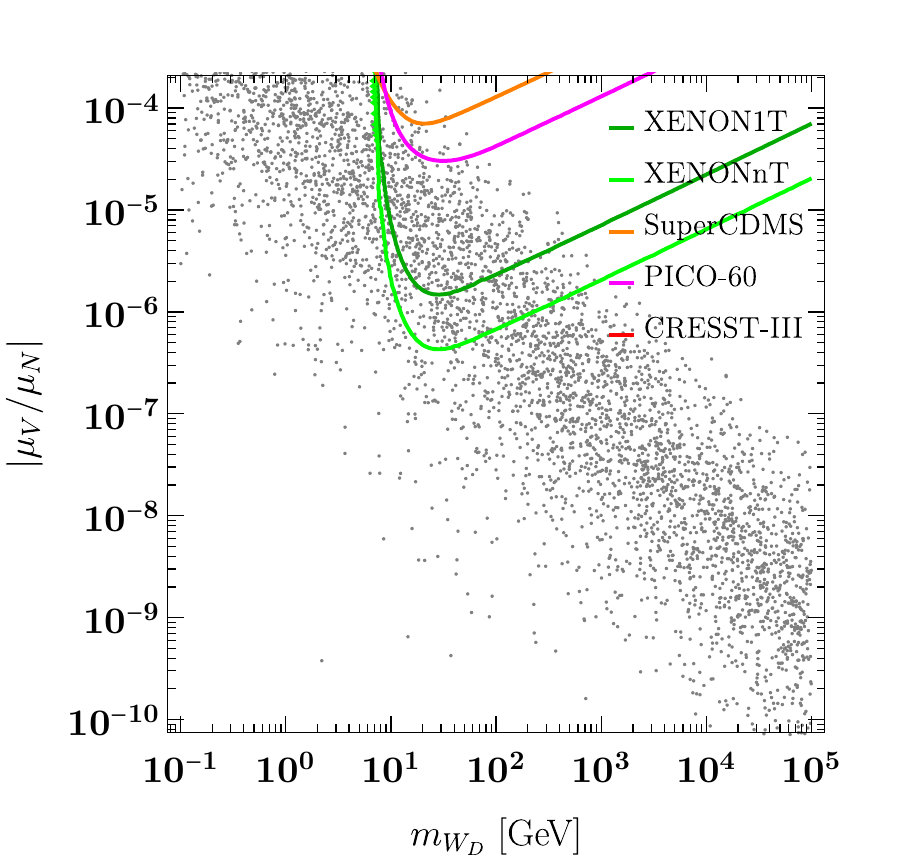}~~~
	\includegraphics[width=6cm,clip]{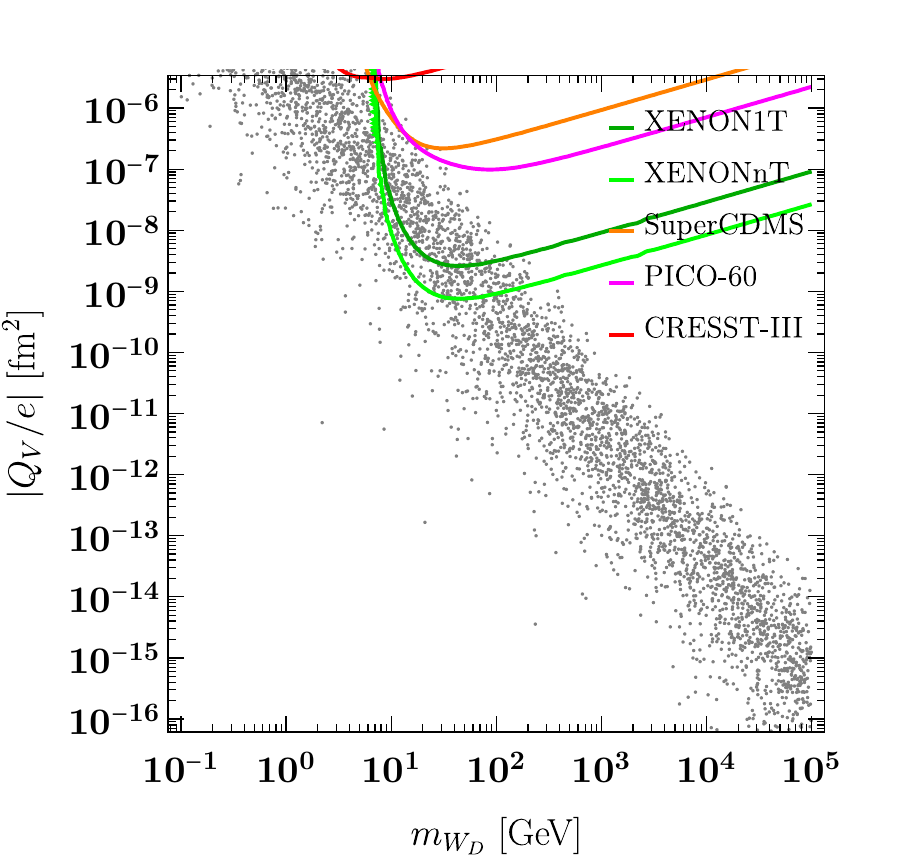}
	\\
	\includegraphics[width=6cm,clip]{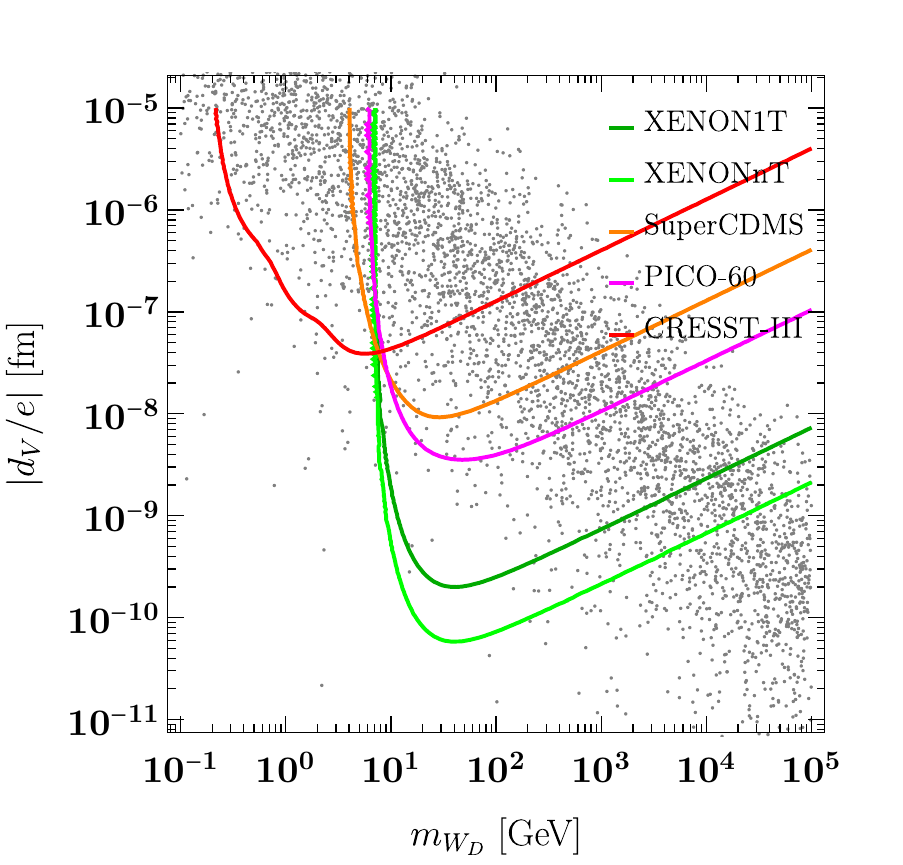}~~~
	\includegraphics[width=6cm,clip]{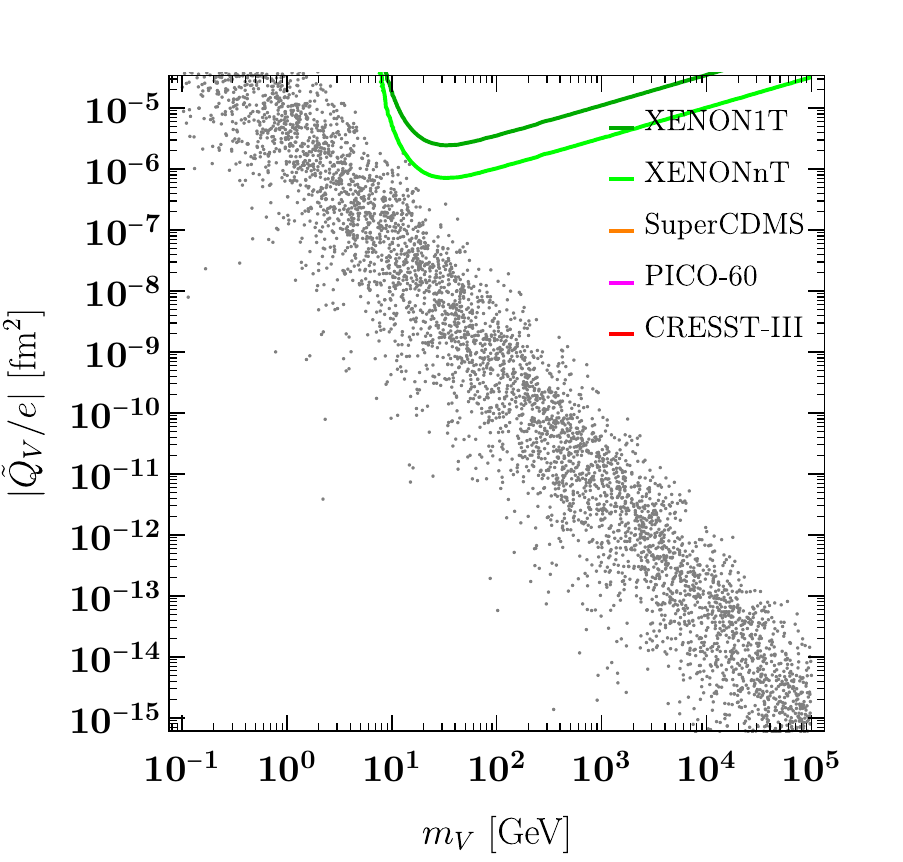}
	\\
	\includegraphics[width=6cm,clip]{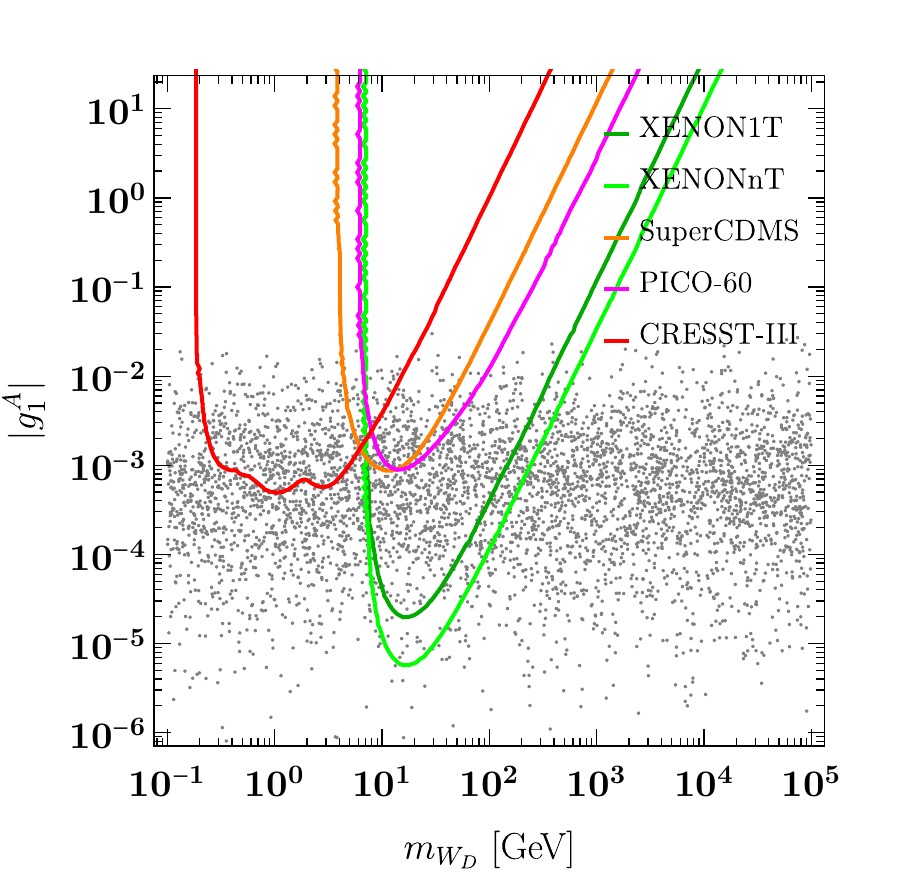}~~~
	\includegraphics[width=6cm,clip]{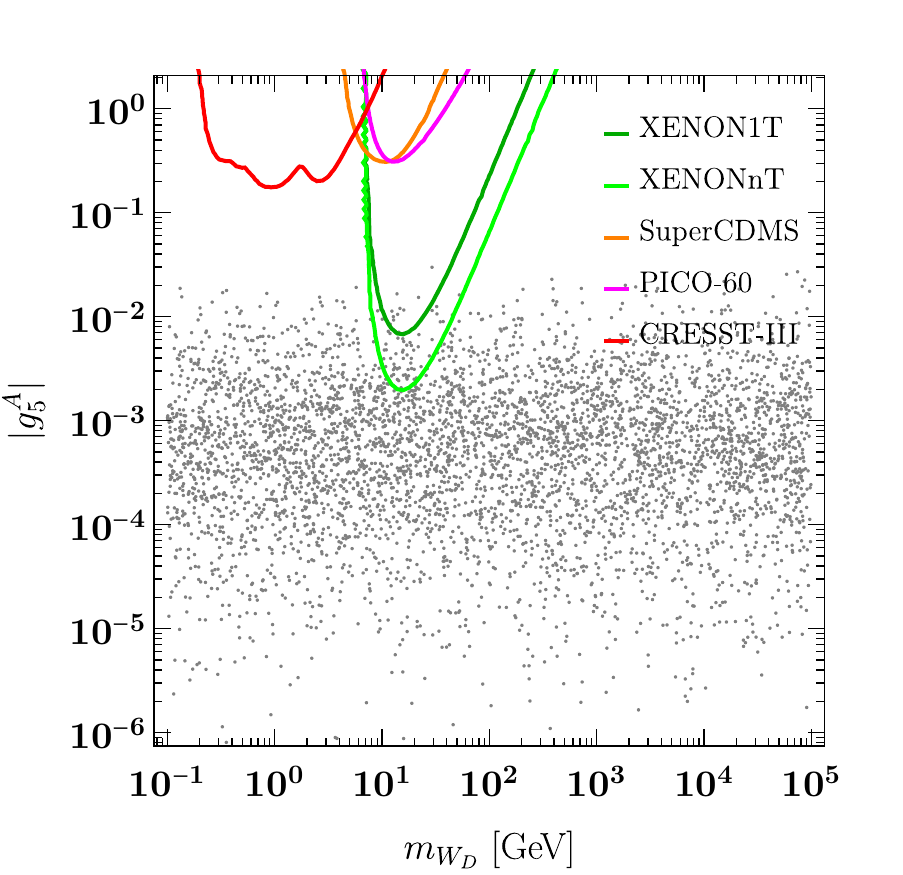}
	\caption{
	Scan plot of the expected electromagnetic multipole moments for the model described in section \ref{sec:Model} (for details see Eq.~(\ref{eq:scan})), confronted to the upper limits from current direct detection experiments as well as the projected sensitivity from XENONnT.}
	\label{fig:formfac-model}
\end{figure}

We show in Figure \ref{fig:formfac-model} a scan plot with the predicted  values of the form factors as a function of the dark matter mass $m_{W_D}$, taking for concreteness $g_D=1$, and the remaining parameters in the ranges:
\begin{align}
&\frac{m_{\Psi_e}}{m_{W_D}}=[1,10]\,,\nonumber\\
&\frac{m_{\Psi_l}}{m_{W_D}}=[1,10]\,,\nonumber\\
&|\lambda_{L,R}|=[0,2]\,,\nonumber\\
&\mbox{Arg}[\lambda_{L,R}]=[0,2\pi]\,.\label{eq:scan}
\end{align}
We also show in the Figure the upper limits on the form factors from various experiments from Figure \ref{fig:DDlimit} (determined assuming that only one form factor contributes to the scattering). Notably, there are portions of the parameter space which can be probed by current experiments, even in the absence of Higgs portal interactions, due especially to the interactions induced by the {\bf CP} violating  moment $d_V$ and by the {\bf CP} conserving moments $\mu_V$, $Q_V$ and $g_1^A$. This is a consequence of the enhancement of the scattering rate induced by these electromagnetic multipoles at low relative velocities, {\it cf.} Eq. ~(\ref{eq:dsigma}),  especially by the electric dipole moment, which is doubly enhanced by $1/E_R$ and by $1/v^2$.

\begin{figure}
	\centering
	\includegraphics[width=7cm,clip]{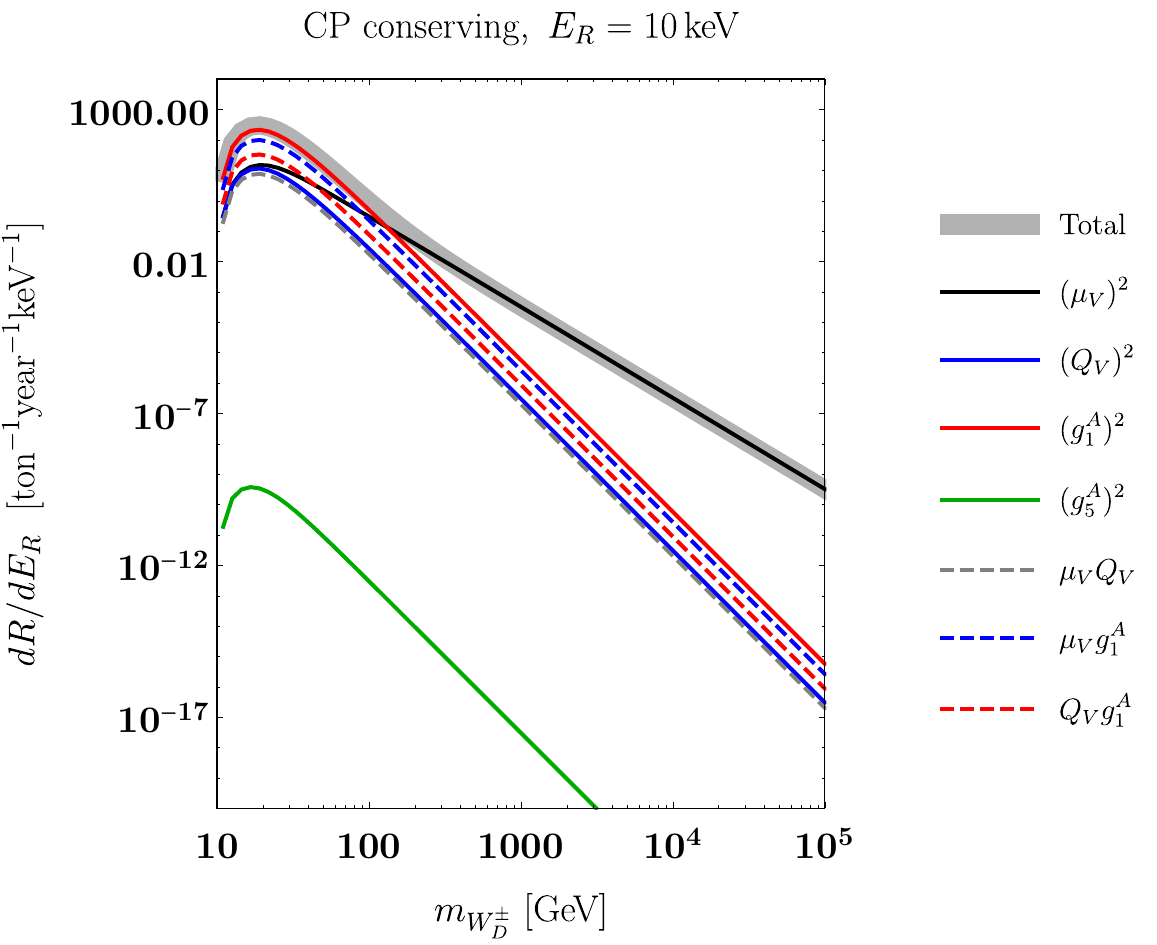}~
	\includegraphics[width=8.5cm,clip]{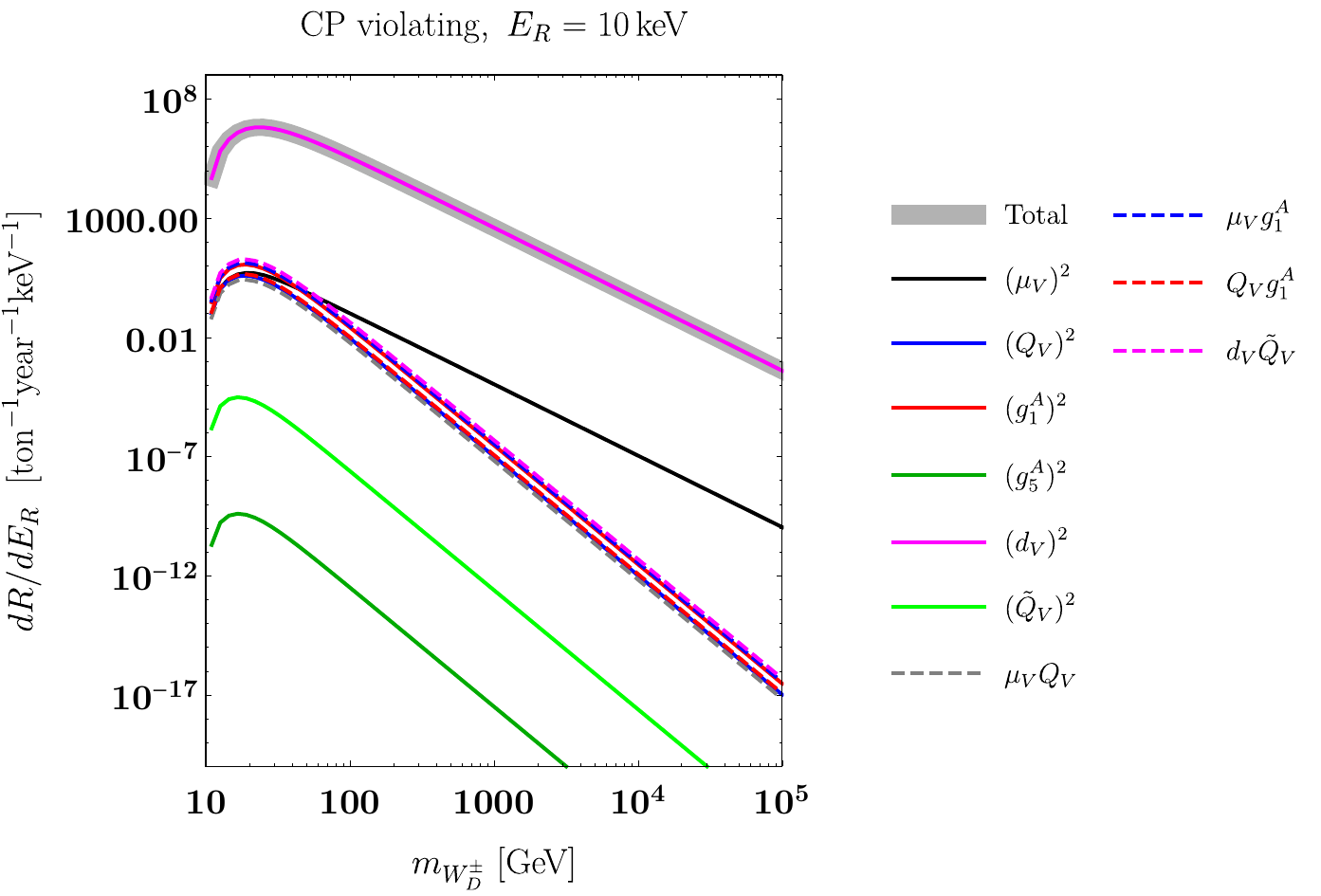}
	\caption{
	Differential event rate for a xenon target at recoil energy  $E_R=10\,\mbox{keV}$  as a function of the dark matter mass, assuming $m_{\Psi_l}=2m_{\Psi_e}=10m_{W_D}$ $g_D=1$, $\lambda_R=-1$ and $\lambda_L=1.5$ ({\bf CP} conserving point, left panel) or $\lambda_L=1.5e^{i\frac{\pi}{3}}$ ({\bf CP} violating point, right panel). }
	\label{fig:dRdE}
\end{figure}

To investigate the relative effect of the different multipoles in the differential event rate, we show in  Figure \ref{fig:dRdE} the contributions of the different terms in Eq. ~(\ref{eq:dsigma}) for a xenon target at recoil energy $E_R=10\,\mbox{keV}$, for some exemplary parameters conserving {\bf CP} (left panel) or violating {\bf CP} (right panel). Concretely, we take $m_{\Psi_l}/m_{W_D}=10$, $m_{\Psi_e}/m_{W_D}=5$, $g_D=1$, $\lambda_R=-1$, as well as $\lambda_L=1.5$ for the {\bf CP} conserving case  and  $\lambda_L=1.5e^{i\frac{\pi}{3}}$ for the {\bf CP} violating case. For the {\bf CP} violating case, the $d_V$ contribution dominates over the whole range of masses analyzed; for the {\bf CP} conserving case, the $\mu_V$ contribution dominates for $m_{W_D}\gtrsim 100\,\mbox{GeV}$, while $g^A_1$ dominates for smaller masses; this is due to the contribution to the interaction vertex from the kinetic mixing.

\begin{figure}
	\centering
	\includegraphics[width=7cm,clip]{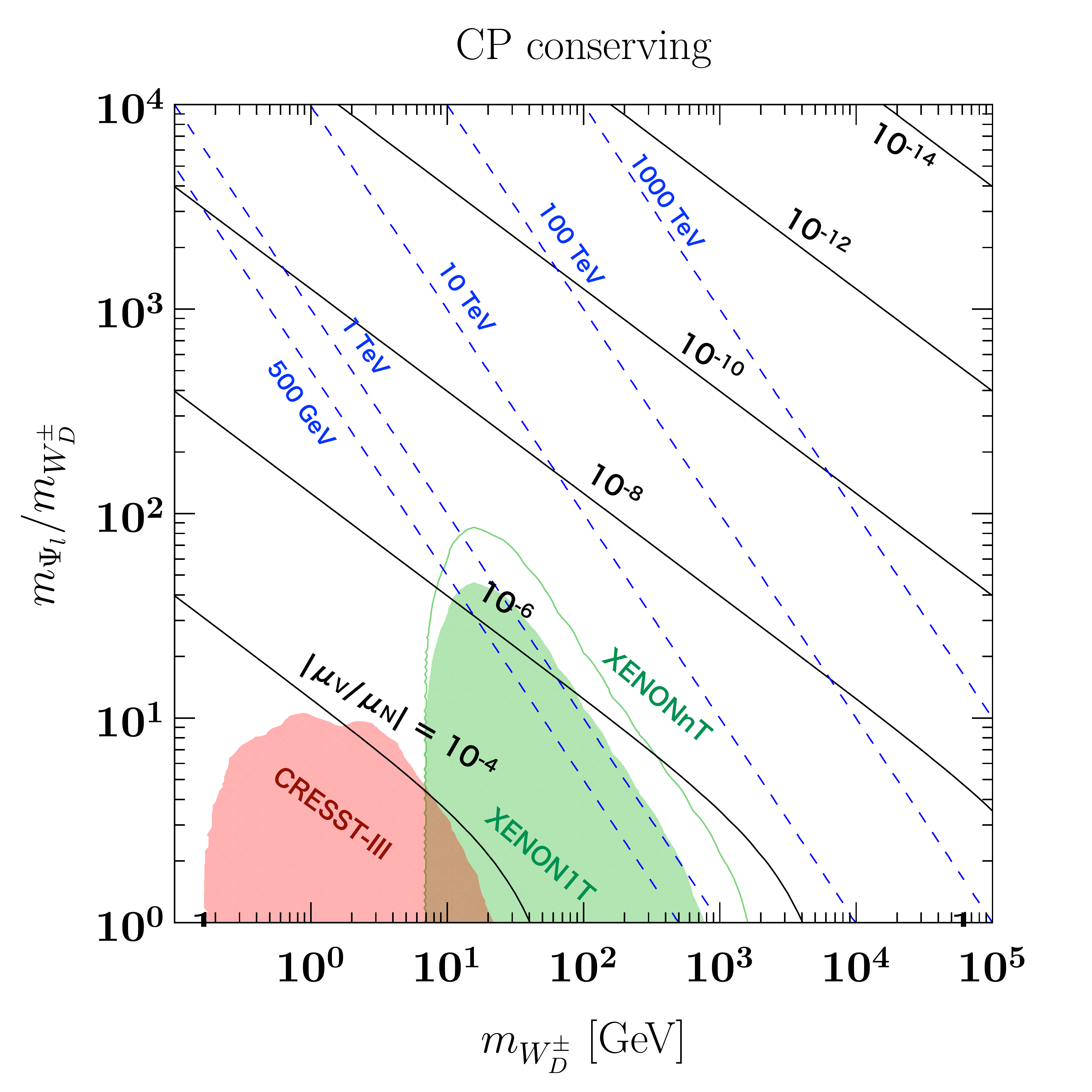}~~~
	\includegraphics[width=7cm,clip]{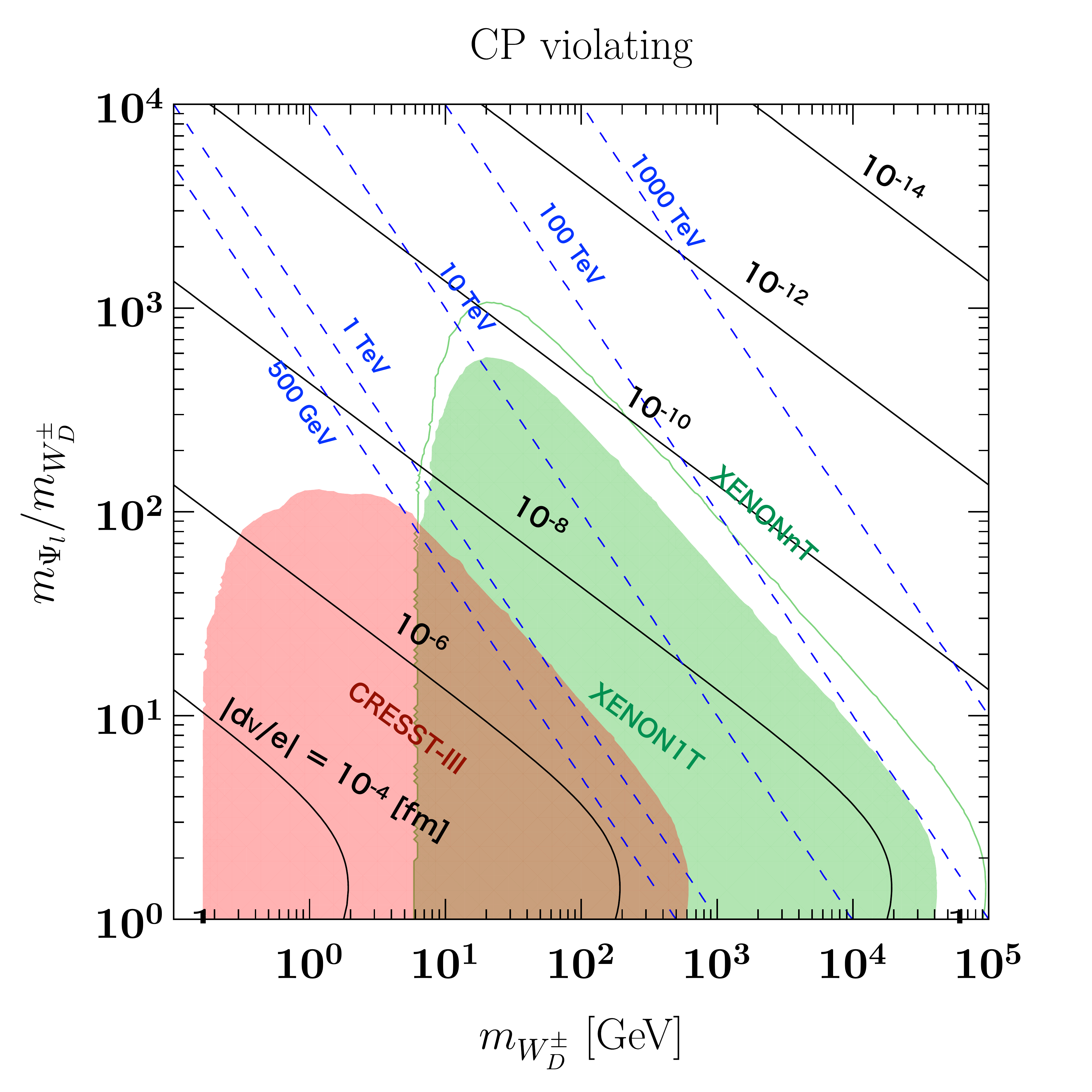}
	\caption{
		Impact of direct detection experiments for the model described in Section \ref{sec:Model} assuming $m_{\Psi_l}=2m_{\Psi_e}$, $g_D=1$, $\lambda_R=-1$, as well as $\lambda_L=1.5$ for the {\bf CP} conserving case (left panel)  and  $\lambda_L=1.5e^{i\frac{\pi}{3}}$ for the {\bf CP} violating case (right panel). The black lines represent  isocontours of the  magnetic dipole moment (left panel) and the electric dipole moment (right panel), while the blue lines represent isocontours of ${\rm min}\{m_N,m_{E^1}\}$. The green and red regions correspond to the $90\,\%$ exclusion limits from the XENON1T and CRESST-III experiments, respectively; the green line corresponds to the future prospect by XENONnT experiment. }
	\label{fig:limit-model}
\end{figure}

Finally, in Figure \ref{fig:limit-model} we investigate the impact of direct detection experiments in probing the parameter space of the model. As before, we fix for concreteness $m_{\Psi_l}=2m_{\Psi_e}$, $g_D=1$, $\lambda_R=-1$, as well as $\lambda_L=1.5$ for the {\bf CP} conserving case (left panel)  and  $\lambda_L=1.5e^{i\frac{\pi}{3}}$ for the {\bf CP} violating case (right panel), and we show as black lines the isocontours of the  magnetic dipole moment ($|\mu_V/\mu_N|$, left panel) and the electric dipole moment ($|d_V/e|\,[\mbox{fm}]$, right panel). 
The blue dashed lines are contours of the smallest mass between $m_{N}$ and $m_{E^1}$. On the other hand, the green and red regions correspond to the $90\,\%$ exclusion limits from the XENON1T and CRESST-III experiments, respectively; the green line corresponds to the future prospect by XENONnT experiment. As shown in the Figure, current experiments probe a significant part of the parameter space, especially for the {\bf CP} violating case, where fermions as heavy as 100 TeV in the loop can induce electric dipole moments at the reach of current experiments. As mentioned in Section 
\ref{sec:TGV} this is due to the double enhancement of the scattering rate mediated by the electric dipole moment by $1/E_R$ and by $1/v^2$.

\section{Summary}
\label{sec:Summary}

We have presented a comprehensive study of electromagnetic multipole moments of the complex vector dark matter candidate, and we have studied their implications for direct detection experiments. We have parametrized the electromagnetic interactions of the vector dark matter by means of seven form factors and we have calculated the differential scattering cross-section of the vector dark matter with the nucleus via the interactions of their multipole moments.

We have set upper limits on the vector dark matter electromagnetic multipole moments from the non-observation of an excess of nuclear recoils in direct detection experiments. For dark matter masses above $\sim 10$ GeV the strongest constraints are set by the XENON1T experiment, and below that mass by the CRESST-III experiment. The strongest limits arise for a dark matter mass  $\simeq 30$ GeV and read $|\mu_V/\mu_N|<2\times 10^{-6}$, $|Q_V/e|<3\times 10^{-9}\,{\rm fm}^2$, $|d_V/e|<2\times 10^{-10}\,{\rm fm}$, $|\tilde Q_V/e|<3\times 10^{-6}\,{\rm fm}^2$, $g_1^A<3\times 10^{-5}$ and $g_5^A<8\times 10^{-3}$.
 
Lastly, we have constructed a concrete model of vector DM where the interactions with the Standard Model are dominated by the electromagnetic multipole moments. The model is based on three ingredients: {\it i)} a ``dark'' non-Abelian gauge symmetry, which is spontaneously broken, {\it ii)} a new $U(1)$ global symmetry, and {\it iii)} new matter particles, charged both under the electromagnetic $U(1)$ symmetry and the ``dark'' non-Abelian symmetry. We find that after the spontaneous breaking of the ``dark'' non-Abelian symmetry, there is a remnant symmetry that stabilizes the vector dark matter against decay. Also, the new matter particles generate via quantum effects the electromagnetic multipole moments for the vector dark matter. We have found that, despite the loop suppression of the multipole moments, the vector dark matter could be detected in current experiments through their electromagnetic  interactions with the nuclei, even when the fermion mass lie in the multi-TeV range.

\section*{Acknowledgments}
We thank Tomohiro Abe for comments on the manuscript.
The work of J.H. and R.N. was supported by JSPS KAKENHI
(Grant Number 20H01895 (J.H.) and 19K14701 (R.N.)) and by Grant-in-Aid for Scientific research
from the Ministry of Education, Science, Sports, and Culture (MEXT),
Japan (Grant Numbers 16H06492 (J.H.) and 18H05542 (R.N.)). The work of J.H. was also supported by JSPS Core-to-Core Program (Grant
Numbers JPJSCCA20200002), and World Premier International Research
Center Initiative (WPI Initiative), MEXT, Japan.
The work of R.N. was also supported by the University of Padua through the ``New Theoretical Tools to Look at the Invisible Universe'' project and by Istituto Nazionale di Fisica Nucleare (INFN) through the ``Theoretical Astroparticle Physics'' (TAsP) project. A.I. would like to thank the KMI for hospitality during the initial stages of this work. 
The work of A.I. has been partially supported by the Deutsche Forschungsgemeinschaft (DFG, German Research Foundation) under Germany's Excellence Strategy – EXC-2094 – 390783311.
\appendix
\section{Effective interactions of an on-shell complex vector  field with the electromagnetic field}
\label{sec:hagiwaraetal}.

In this appendix we review the general structure of the  $VV^\dagger\gamma$ vertex derived in \cite{Hagiwara:1986vm}, where $V$ is a complex vector field, and we present the effective interaction Lagrangian of an on-shell vector field $V^\mu$ with the electromagnetic field $A^\mu$, keeping terms up to dimension six. The general interaction vertex reads:
\begin{align}
\Gamma^{\alpha\beta\mu}_{VV^\dag\gamma}(q,\bar{q},p)/e
&=
f^A_1(p^2) Q^\mu g^{\alpha\beta}
-\frac{f^A_2(p^2)}{m^2_{V}}Q^\mu p^\alpha p^\beta
+f^A_3(p^2)(p^\alpha g^{\mu\beta}-p^\beta g^{\mu\alpha})\nn\\
&
+{if^A_4(p^2)}
(p^\alpha g^{\mu\beta}+p^\beta g^{\mu\alpha})
+{if^A_5(p^2)}\epsilon^{\mu\alpha\beta\rho}Q_\rho\nn\\
&
-f^A_6(p^2)\epsilon^{\mu\alpha\beta\rho}p_\rho
-\frac{f^A_7(p^2)}{m^2_{V}}Q^\mu \epsilon^{\alpha\beta \rho\sigma} p_\rho Q_\sigma\,,
\label{eq:VVA}
\end{align}
where $q^\mu$, $\bar q^\mu$ and $p^\mu$ are, respectively, the 4-momenta of the fields $V$, $V^\dagger$ and $A$, and $Q^\mu=q^\mu-\bar{q}^\mu$. Here, we have imposed the conditions
\begin{eqnarray}
  \partial^\mu V_\mu=0, &&\partial^\mu A_\mu=0.
  \nonumber
\end{eqnarray}
The first condition is justified for on-shell $V_\mu$, since it is derived from the equation of motion, which for (free) $V_\mu$ is reads
$(\Box+m^2_V)V_\nu =\partial_\nu(\partial^\mu V_\mu)$. The second condition is justified because the scalar component does not contribute to the scattering amplitude.

The form factors are regular at $p^2=0$.  Let us note that $f^A_{4,5}(0)=0$  due to the $U(1)_{\rm EM}$ gauge invariance. Further, $f_1^A(0)$ corresponds to the electromagnetic charge of $V$, and it is zero in our case. $f^A_i(p^2)~(i=1,2,\cdots,7)$ are expanded around $p^2\simeq 0$ as 
\begin{align}
f^A_1(p^2)&=\frac{p^2}{2m_V^2}(g^A_1+\lambda_A)+{\cal O}(p^4), \nonumber\\
f^A_2(p^2)&=\lambda_A +{\cal O}(p^2), \nonumber\\
f^A_3(p^2)&= \kappa_A+\lambda_A +{\cal O}(p^2),\nonumber\\
f^A_4(p^2)&=\frac{p^2}{m_V^2}g^A_4 +{\cal O}(p^4), \nonumber\\
f^A_5(p^2)&=\frac{p^2}{m_V^2}g^A_5+{\cal O}(p^4), \nonumber\\
f^A_6(p^2)&=\tilde\kappa_A-\tilde\lambda_A +{\cal O}(p^2), \nonumber\\
f^A_7(p^2)&=-\frac{\tilde\lambda_A}{2}+{\cal O}(p^2).
\end{align}
The coefficients $g_1^A$, $g_4^A$, $g_5^A$,
$\lambda_A$, $\tilde\kappa_A$, $\tilde\lambda_A$, and $\kappa_A$ parametrize the strength of the effective interactions of an on-shell vector $V^\mu$ with the electromagnetic field $A^\mu$, as defined in Eq.~(\ref{effective_lag}).

\section{Direct detection event rate}
\label{sec:DDrate}
We summarize in this Appendix how we calculated the expected number of signal events in the experiments employed in our analysis. 
\subsection{XENON1T/nT}
\label{sec:XENON1T}
The XENON collaboration uses a liquid xenon detector with a dual-phase time projection chamber (TPC). The signal from nucleon recoils can be efficiently discriminated from the background signals from the ratio between the primary ($S_1$) and secondary  ($S_2$) scintillation light signals. The scintillation light is converted into photoelectrons (PE) by the photomultiplier tubes (PMT). 

The signal rate in number of photoelectrons $n$ can be calculated from \cite{Aprile:2011hx};
\begin{align}
 \frac{dR}{dn} = 
 \int_{E^{\rm{max}}_R}^{E^{\rm{min}}_R} dE_R \, \epsilon (E_R) \,
\text{Poiss} (n|\nu (E_R)) \frac{dR}{dE_R}
~,
\end{align}
where 
${dR}/{dE_R}$ denotes the differential event rate for the dark matter scattering off a $^{131}$Xe nucleus, while
$E^{\rm{min}}_R=4.9$ keV and $E^{\rm{max}}_R=40.9$ keV. Further, 
$\epsilon (E_R)$ is the detection efficiency, which we take from Fig.~1 of Ref.~\cite{Aprile:2018dbl}. Finally, $\nu (E_R)$ is the expected number of PEs for a given
recoil energy $E_R$, which we obtain from the S1
yield given in the lower left panel of Fig.~13 of Ref.~\cite{Aprile:2015uzo}.
In our analysis, we focus on the central detector region with a mass of 0.65\,t and consider only events between the median of the nuclear recoil band and the lower $2\sigma$ quantile. This approach reduces the background level, as argued in Refs.~\cite{Athron:2018hpc, Athron:2018ipf}. We therefore multiply the detection efficiency by an additional factor 0.475 to take into account our reference region.
Next, we determine the differential event rate as the function of the primary scintillation light yield. The differential event rate reads:
\begin{align}
 \frac{dR}{dS_1} = 
\sum_{n=1}^{\infty} \text{Gauss} (S_1|n, \sqrt{n} \sigma_{\text{PMT}})
\frac{dR}{dn}
~.
\end{align}
Here, $\sigma_{\text{PMT}}$ is the average single-PE
resolution of the photomultipliers, for which we conservatively take $\sigma_{\text{PMT}} = 0.4$ \cite{Aprile:2015lha, Barrow:2016doe}.
Finally the expected number of events in the energy bin $[S^{\rm{min}}_1,S^{\rm{max}}_1]$ is obtained from
\begin{align}
N_{\rm{th}}[S^{\rm{min}}_1,S^{\rm{max}}_1]
=
w_{\rm{exp}}\int_{S^{\rm{min}}_1}^{S^{\rm{max}}_1}
\frac{dR}{dS_1}\,.
\end{align}
where $w_{\rm exp}$ is the exposure of the experiment.

Following Refs.~\cite{Athron:2018hpc, Athron:2018ipf}, we divide the signal region into two parts, which correspond to $S_1\in [3,35]$\,PE and $S_2\in [35,70]$\,PE, respectively. For each energy bin, we calculate the Test Statistic (TS) function, defined as (see {\it e.g.} \cite{DelNobile:2013sia}):
\begin{equation}
 \text{TS} = -2 \ln \biggl[\frac{{\cal L}(N_{\text{th}})}{{\cal
  L}_{\text{bkg}}}\biggr]~,
  \label{eq:TS}
\end{equation}
with
\begin{equation}
 {\cal L} (N_{\text{th}}) = \frac{1}{N_{\text{obs}}!} 
(N_{\text{th}} +N_{\text{bkg}})^{N_{\text{obs}}} \exp \bigl\{
- (N_{\text{th}} +N_{\text{bkg}})
\bigr\} ~,
\end{equation}
and ${\cal L}_{\text{bkg}} \equiv {\cal L} (0)$. $N_{\text{obs}}$ and $N_{\text{bkg}}$ are the numbers of the observed and background events, respectively.

The XENON1T collaboration has reported in \cite{Aprile:2018dbl} the latest results of their search, using an exposure $w_{\text{exp}} = 278.8~ \text{days}\times 1.30(1)$~ton. Following Refs.~\cite{Athron:2018hpc, Athron:2018ipf}, we adopt $(N_{\text{obs}},N_{\text{bkg}})=(0,0.46)$ for the first energy bin and $(N_{\text{obs}},N_{\text{bkg}})=(2,0.34)$ for the second energy bin. Finally, we derive the 90\% CL upper limit on the number of signal events by requiring $\text{TS} > 2.71$ in each energy bin.

To estimate the future prospect of XENONnT with an exposure $w_{\text{exp}} = 20\,\text{t} \cdot \text{yrs}$ \cite{Aprile:2015uzo}, we follow  Ref.~\cite{Witte:2017qsy} and we apply the maximum gap method \cite{Yellin:2002xd} under the assumption of zero observed events. Namely, we require
$1- \exp(-N_{\text{th}}) \geq 0.9$, which corresponds to
$N_{\text{th}} \lesssim 2.3$. 
\subsection{SuperCDMS}
\label{sec:SuperCDMS}
The SuperCDMS detector consists of 15 Ge target crystals, each instrumented with ionization and phonon detectors. The measured ionization and phonon energies are used to derive the recoil energy and the ionization yield. The information from the ionization yield can be used to distinguish signal from background.

Following Ref.~\cite{DelNobile:2013sia}, we estimate the DM event rate as
\begin{align}
N_{\rm{th}}
=
w_{\rm{exp}}
\int^{E_{\rm{max}}}_{E_{\rm{min}}}dE_R\,,
\epsilon(E_R)\frac{dR}{dE_R}\,,
\end{align}
where ${dR}/{dE_R}$ denotes the differential scattering rate of dark matter particles off a $^{73}$Ge nucleus, $E_{\rm{min}}=1.6$\,keV, $E_{\rm{max}}=10$\,keV,  $w_{\rm{exp}}$ is the exposure, and $\epsilon(E_R)$ is the efficiency, which we take from Fig. 1 of \cite{Agnese:2014aze}.

The SuperCDMs collaboration presented in \cite{Agnese:2014aze} the results of their first search, based on an exposure 
$w_{\rm{exp}}=577 \,{\rm kg}\cdot {\rm days}$, reporting  $N_{\rm{obs}}=11$. On the other hand, the backgrounds in their experiment are not fuly understood, therefore we conservatively take $N_{\rm{bkg}}=0$ in the derivation of the upper limit of signal events. The $90\%$ C.L. limit corresponds to $N_{\rm{th}}<16.6$ \cite{Ferrer:2015bta}.
\subsection{CRESST-III}
\label{sec:CRESST-III}
The CRESST-III experiment employs a CaWO$_4$ crystal target as cryogenic calorimeters. The discrimination of the dark matter signal from the background is performed by measuring simultaneously the phonon/heat and the scintillation light signals. The design aims to achieve a low threshold for the recoil energy, smaller than $100\,\mbox{eV}$.

The expected event rate in the energy bin $[E_{\rm{min}},E_{\rm{max}}]$, can be calculated from
\begin{align}
N_{\rm{th}}
=
w_{\rm{exp}}
\int^{E_{\rm{max}}}_{E_{\rm{min}}}\sum_{i=\{\rm{Ca},\rm{O},\rm{W}\}}
f_i
\epsilon_i(E_R)
\frac{dR_i}{dE_R}\,.
\end{align}
Here $dR_i/dE_R$ denotes the differential cross section for the DM scattering off the nucleus $i$=Ca, O, W, and $w_{\rm exp}$ is the exposure. Further, $\epsilon_i(E_R)$ is the detector efficiency for the nucleus $i$, which we read from the data implemented in {\tt{DDCalc-2.0.0}} \cite{DDCalc:url,DarkBit:2017lvb}. Finally, $f_i$ denotes the mass fraction for element $i$: $f_{\rm{Ca}}=0.1392$, $f_{\rm{O}}=0.22228$, and $f_{\rm{W}}=0.63852$, respectively.

In the first run (from $05/2016-02/2018$), five detectors reached/exceeded the design goal \cite{Mancuso:2018zoh}. Among the five detectors, the detector called ``detector A'' achieved the lowest energy threshold $\simeq 30\,\mbox{eV}$ \cite{Abdelhameed:2019hmk}.  The results from the detector A give the largest sensitivity to low mass dark matter candidates. The total exposure was $w_{\rm{exp}}=5.689\,\mbox{kg}\times \,\mbox{days}$ \cite{Abdelhameed:2019mac}.

To derive the 90\%\,C.L. limit, we consider 10 energy bins of uniform size in log-scale between $30$\,eV and $1$\,keV, and simply assume that the number of signal events follows a Poisson distribution with 
$N_{\rm{obs}}=\{355,219,103,57, 21, 7, 9, 11, 18, 74\}$. To determine the $90\,\%$ C.L. limit on the event rate we 
require that the Poisson likelihoods $\mathcal{L}(N_{\rm{th}})$  satisfy 
\begin{align}
\chi(N_{\rm{th}})
=
-2\ln\mathcal{L}(N_{\rm{th}})
+2\ln\mathcal{L}(N_{\rm{obs}})
<
2.71\,
\end{align}
in each of the bins, where the Poisson likelihood is calculated from
\begin{align}
-2\ln\mathcal{L}(N_{\rm{th}})
=
2N_{\rm{th}}-N_{\rm{obs}}+N_{\rm{obs}}\ln\frac{N_{\rm{obs}}}{N_{\rm{th}}}\,.
\end{align}
\subsection{PICO-60}
\label{sec:PICO-60}
The PICO-60 collaboration employs a C$_3$F$_8$ superheated liquid detector. The expected event number can be calculated from 
\begin{align}
N_{\rm{th}}
=
w_{\rm{exp}}
\sum_{i=\{\rm{C},\rm{F}\}}
\int_0^\infty dE_R
f_i
\mathcal{P}_i(E_R)\frac{dR_i}{dE_R}\,,
\end{align}
Here $dR_i/dE_R$ denotes the differential cross section for the DM scattering off the nucleus $i$=C, F, and $w_{\rm{exp}}$ is the exposure. Further, $\mathcal{P}(E_R)$ is the bubble nucleon efficiency for given a recoil energy $E_R$. We read the efficiency for $\rm{F}$ and $\rm{C}$ from Figure 3 of Ref.~\cite{Amole:2019fdf}. Finally, $f_i$ denotes the mass fraction for element $i$: $f_{\rm{C}}=0.19164$ and $f_{\rm{F}}=0.80836$, respectively.

The results of the dark matter search were reported in Ref.~\cite{Amole:2019fdf} for an exposure $w_{\rm{exp}}=48.9\,\mbox{kg}\times 52.6$\,days. The collaboration reported the observation of 3 candidate events (based on the single bubble selection) while the number of background events is determined to be $1.0\pm 0.4$. We estimate the 90\% C.L. bound on $N_{\rm{th}}$ by inserting $N_{\rm{obs}}=3$ and $N_{\rm{bkg}}=1$ into Eq.~(\ref{eq:TS}) and by requiring that $\mbox{TS}>2.71$. We  obtain $N_{\rm{th}}\lesssim 7.92$. as 90\% C.L. bound on the event rate.

\section{Analytic expressions for the electromagnetic form factors}
\label{app:form}

The loop function $\mathcal{G}$ are given by:\footnote{
We used {\tt{Package-X}} \cite{Patel:2015tea} to evaluate the one-loop diagrams.}
\begin{align}
\mathcal{G}^{(1)}_\mu(x,y)
&=-2
-2{\lambda} 
\Big((x^2-y^2)^2 -(x^2+y^2)
\Big)
+(x^2-y^2)\log \left(\frac{{x^2}}{{y^2}}\right)
\,,\\
\mathcal{G}^{(2)}_\mu(x,y)
&=  
4 x y\lambda(x,y)-\frac{2 x y }{{x^2}-{y^2}} \log
\left(\frac{{x^2}}{{y^2}}\right)
\,,\\
\mathcal{G}^{(1)}_Q(x,y)
&=
-\frac{4}{3}\Big(\lambda(x,y) 
\Big(x^4-x^2(2y^2+3)+y^4-3y^2+2\Big)+1
\Big)
\nn\\
&
+\frac{2\Big(x^4-2x^2(y^2+1)+y^2(y^2-2)\Big)}{3(x^2-y^2)}\log
\left(\frac{{x^2}}{{y^2}}\right)
\,,\\
\mathcal{G}^{(2)}_Q(x,y)
&=
-8 x y  \lambda(x,y) +\frac{4 x y }{{x^2}-{y^2}}\log
\left(\frac{{x^2}}{{y^2}}\right)
\,,\\
\mathcal{G}^{(1)}_1(x,y)
&=
\frac{4}{3}\frac{\lambda(x,y)}{\kappa(x,y)}\Big(3+4(x^2-y^2)^4+(x^2-y^2)^2(8-13(x^2+y^2))-11(x^2+y^2)+9(x^2+y^2)^2\Big)\nonumber\\
&+\frac{4}{3}\frac{\Big(4(x^2-y^2)^2-7(x^2+y^2)+3\Big)}{\kappa(x,y)}-\frac{2}{3}\Big(\frac{4(x^2-y^2)^2-(x^2+y^2)+2}{x^2-y^2}\Big)\log\Big(\frac{x^2}{y^2}\Big)    
\,,\\
\mathcal{G}^{(2)}_1(x,y)
&= 
-\frac{8}{3}\frac{ xy}{\kappa(x,y)} -\frac{8}{3}\frac{\lambda(x,y)}{\kappa(x,y)} x y
\Big((x^2-y^2)^2-3(x^2+y^2)+2\Big)
\nn\\
&+\frac{4}{3}\Big(\frac{ xy}{x^2-y^2}\Big) \log \left(\frac{x^2}{y^2}\right)
\,,\\
\mathcal{G}^{(1)}_d(x,y)
&=
8 \,x\, y\,\lambda(x,y)
\,,\\
\mathcal{G}_{\tilde{Q}}(x,y)
&=
-16 \,x\,y\,\lambda(x,y)
\,,
\\
\mathcal{G}_5(x,y)
&=
\lambda(x,y) \left(-4 x^4+x^2 \left(8 y^2+4\right)-4 y^4+4 y^2-\frac{8}{3}\right)-2 \left(x^2-y^2\right) \log \left(\frac{y^2}{x^2}\right)-4
\,,
\end{align}
where $\lambda$ and $\kappa$ are defined as
\begin{align}
\kappa(x,y)
&
=(1-x^2-y^2)^2-4x^2y^2\,,\\
\lambda(x,y)
&=
\frac{1}{\kappa^{1/2}(x,y)}
\log\l(
\frac{\kappa^{1/2}(x,y)-(1-x^2-y^2)}{2xy}
\r)\,.
\end{align}

\bibliographystyle{JHEP}
\bibliography{papers}

\providecommand{\href}[2]{#2}\begingroup\raggedright\begin{thebibliography}{10}

\bibitem{Bertone:2004pz}
G.~Bertone, D.~Hooper and J.~Silk, \emph{{Particle dark matter: Evidence,
  candidates and constraints}},
  \href{https://doi.org/10.1016/j.physrep.2004.08.031}{\emph{Phys. Rept.}
  {\bfseries 405} (2005) 279}
  [\href{https://arxiv.org/abs/hep-ph/0404175}{{\ttfamily hep-ph/0404175}}].

\bibitem{Bergstrom:2000pn}
L.~Bergström, \emph{{Nonbaryonic dark matter: Observational evidence and
  detection methods}},
  \href{https://doi.org/10.1088/0034-4885/63/5/2r3}{\emph{Rept. Prog. Phys.}
  {\bfseries 63} (2000) 793}
  [\href{https://arxiv.org/abs/hep-ph/0002126}{{\ttfamily hep-ph/0002126}}].

\bibitem{Bertone:2010zza}
J.~Silk et~al., \emph{{Particle Dark Matter: Observations, Models and
  Searches}}. Cambridge Univ. Press, Cambridge, 2010,
  \href{https://doi.org/10.1017/CBO9780511770739}{10.1017/CBO9780511770739}.

\bibitem{Holdom:1985ag}
B.~Holdom, \emph{{Two U(1)'s and Epsilon Charge Shifts}},
  \href{https://doi.org/10.1016/0370-2693(86)91377-8}{\emph{Phys. Lett. B}
  {\bfseries 166} (1986) 196}.

\bibitem{Davidson:2000hf}
S.~Davidson, S.~Hannestad and G.~Raffelt, \emph{{Updated bounds on millicharged
  particles}}, \href{https://doi.org/10.1088/1126-6708/2000/05/003}{\emph{JHEP}
  {\bfseries 05} (2000) 003}
  [\href{https://arxiv.org/abs/hep-ph/0001179}{{\ttfamily hep-ph/0001179}}].

\bibitem{Bagnasco:1993st}
J.~Bagnasco, M.~Dine and S.~D. Thomas, \emph{{Detecting technibaryon dark
  matter}}, \href{https://doi.org/10.1016/0370-2693(94)90830-3}{\emph{Phys.
  Lett. B} {\bfseries 320} (1994) 99}
  [\href{https://arxiv.org/abs/hep-ph/9310290}{{\ttfamily hep-ph/9310290}}].

\bibitem{Pospelov:2000bq}
M.~Pospelov and T.~ter Veldhuis, \emph{{Direct and indirect limits on the
  electromagnetic form-factors of WIMPs}},
  \href{https://doi.org/10.1016/S0370-2693(00)00358-0}{\emph{Phys. Lett. B}
  {\bfseries 480} (2000) 181}
  [\href{https://arxiv.org/abs/hep-ph/0003010}{{\ttfamily hep-ph/0003010}}].

\bibitem{Sigurdson:2004zp}
K.~Sigurdson, M.~Doran, A.~Kurylov, R.~R. Caldwell and M.~Kamionkowski,
  \emph{{Dark-matter electric and magnetic dipole moments}},
  \href{https://doi.org/10.1103/PhysRevD.70.083501}{\emph{Phys. Rev. D}
  {\bfseries 70} (2004) 083501}
  [\href{https://arxiv.org/abs/astro-ph/0406355}{{\ttfamily
  astro-ph/0406355}}].

\bibitem{Masso:2009mu}
E.~Masso, S.~Mohanty and S.~Rao, \emph{{Dipolar Dark Matter}},
  \href{https://doi.org/10.1103/PhysRevD.80.036009}{\emph{Phys. Rev. D}
  {\bfseries 80} (2009) 036009}
  [\href{https://arxiv.org/abs/0906.1979}{{\ttfamily 0906.1979}}].

\bibitem{Barger:2010gv}
V.~Barger, W.-Y. Keung and D.~Marfatia, \emph{{Electromagnetic properties of
  dark matter: Dipole moments and charge form factor}},
  \href{https://doi.org/10.1016/j.physletb.2010.12.008}{\emph{Phys. Lett. B}
  {\bfseries 696} (2011) 74} [\href{https://arxiv.org/abs/1007.4345}{{\ttfamily
  1007.4345}}].

\bibitem{Banks:2010eh}
T.~Banks, J.-F. Fortin and S.~Thomas, \emph{{Direct Detection of Dark Matter
  Electromagnetic Dipole Moments}},
  \href{https://arxiv.org/abs/1007.5515}{{\ttfamily 1007.5515}}.

\bibitem{Weiner:2012gm}
N.~Weiner and I.~Yavin, \emph{{UV completions of magnetic inelastic and
  Rayleigh dark matter for the Fermi Line(s)}},
  \href{https://doi.org/10.1103/PhysRevD.87.023523}{\emph{Phys. Rev. D}
  {\bfseries 87} (2013) 023523}
  [\href{https://arxiv.org/abs/1209.1093}{{\ttfamily 1209.1093}}].

\bibitem{Fukushima:2013efa}
K.~Fukushima and J.~Kumar, \emph{{Dipole Moment Bounds on Dark Matter
  Annihilation}}, \href{https://doi.org/10.1103/PhysRevD.88.056017}{\emph{Phys.
  Rev. D} {\bfseries 88} (2013) 056017}
  [\href{https://arxiv.org/abs/1307.7120}{{\ttfamily 1307.7120}}].

\bibitem{Kopp:2014tsa}
J.~Kopp, L.~Michaels and J.~Smirnov, \emph{{Loopy Constraints on Leptophilic
  Dark Matter and Internal Bremsstrahlung}},
  \href{https://doi.org/10.1088/1475-7516/2014/04/022}{\emph{JCAP} {\bfseries
  04} (2014) 022} [\href{https://arxiv.org/abs/1401.6457}{{\ttfamily
  1401.6457}}].

\bibitem{Ibarra:2015fqa}
A.~Ibarra and S.~Wild, \emph{{Dirac dark matter with a charged mediator: a
  comprehensive one-loop analysis of the direct detection phenomenology}},
  \href{https://doi.org/10.1088/1475-7516/2015/05/047}{\emph{JCAP} {\bfseries
  1505} (2015) 047} [\href{https://arxiv.org/abs/1503.03382}{{\ttfamily
  1503.03382}}].

\bibitem{Primulando:2015lfa}
R.~Primulando, E.~Salvioni and Y.~Tsai, \emph{{The Dark Penguin Shines Light at
  Colliders}}, \href{https://doi.org/10.1007/JHEP07(2015)031}{\emph{JHEP}
  {\bfseries 07} (2015) 031}
  [\href{https://arxiv.org/abs/1503.04204}{{\ttfamily 1503.04204}}].

\bibitem{Sandick:2016zut}
P.~Sandick, K.~Sinha and F.~Teng, \emph{{Simplified Dark Matter Models with
  Charged Mediators: Prospects for Direct Detection}},
  \href{https://doi.org/10.1007/JHEP10(2016)018}{\emph{JHEP} {\bfseries 10}
  (2016) 018} [\href{https://arxiv.org/abs/1608.00642}{{\ttfamily
  1608.00642}}].

\bibitem{Herrero-Garcia:2018koq}
J.~Herrero-Garcia, E.~Molinaro and M.~A. Schmidt, \emph{{Dark matter direct
  detection of a fermionic singlet at one loop}},
  \href{https://doi.org/10.1140/epjc/s10052-018-5935-5}{\emph{Eur. Phys. J. C}
  {\bfseries 78} (2018) 471}
  [\href{https://arxiv.org/abs/1803.05660}{{\ttfamily 1803.05660}}].

\bibitem{Hisano:2018bpz}
J.~Hisano, R.~Nagai and N.~Nagata, \emph{{Singlet Dirac Fermion Dark Matter
  with Mediators at Loop}},
  \href{https://doi.org/10.1007/JHEP12(2018)059}{\emph{JHEP} {\bfseries 12}
  (2018) 059} [\href{https://arxiv.org/abs/1808.06301}{{\ttfamily
  1808.06301}}].

\bibitem{Kayser:1983wm}
B.~Kayser and A.~S. Goldhaber, \emph{{{CPT} and {CP} Properties of Majorana
  Particles, and the Consequences}},
  \href{https://doi.org/10.1103/PhysRevD.28.2341}{\emph{Phys. Rev. D}
  {\bfseries 28} (1983) 2341}.

\bibitem{Radescu:1985wf}
E.~Radescu, \emph{{Comments on the Electromagnetic Properties of Majorana
  Fermions}}, \href{https://doi.org/10.1103/PhysRevD.32.1266}{\emph{Phys. Rev.
  D} {\bfseries 32} (1985) 1266}.

\bibitem{Servant:2002aq}
G.~Servant and T.~M. Tait, \emph{{Is the lightest Kaluza-Klein particle a
  viable dark matter candidate?}},
  \href{https://doi.org/10.1016/S0550-3213(02)01012-X}{\emph{Nucl. Phys. B}
  {\bfseries 650} (2003) 391}
  [\href{https://arxiv.org/abs/hep-ph/0206071}{{\ttfamily hep-ph/0206071}}].

\bibitem{Cheng:2002ej}
H.-C. Cheng, J.~L. Feng and K.~T. Matchev, \emph{{Kaluza-Klein dark matter}},
  \href{https://doi.org/10.1103/PhysRevLett.89.211301}{\emph{Phys. Rev. Lett.}
  {\bfseries 89} (2002) 211301}
  [\href{https://arxiv.org/abs/hep-ph/0207125}{{\ttfamily hep-ph/0207125}}].

\bibitem{Hubisz:2004ft}
J.~Hubisz and P.~Meade, \emph{{Phenomenology of the littlest Higgs with
  T-parity}}, \href{https://doi.org/10.1103/PhysRevD.71.035016}{\emph{Phys.
  Rev. D} {\bfseries 71} (2005) 035016}
  [\href{https://arxiv.org/abs/hep-ph/0411264}{{\ttfamily hep-ph/0411264}}].

\bibitem{Birkedal:2006fz}
A.~Birkedal, A.~Noble, M.~Perelstein and A.~Spray, \emph{{Little Higgs dark
  matter}}, \href{https://doi.org/10.1103/PhysRevD.74.035002}{\emph{Phys. Rev.
  D} {\bfseries 74} (2006) 035002}
  [\href{https://arxiv.org/abs/hep-ph/0603077}{{\ttfamily hep-ph/0603077}}].

\bibitem{Hambye:2008bq}
T.~Hambye, \emph{{Hidden vector dark matter}},
  \href{https://doi.org/10.1088/1126-6708/2009/01/028}{\emph{JHEP} {\bfseries
  01} (2009) 028} [\href{https://arxiv.org/abs/0811.0172}{{\ttfamily
  0811.0172}}].

\bibitem{Hisano:2010yh}
J.~Hisano, K.~Ishiwata, N.~Nagata and M.~Yamanaka, \emph{{Direct Detection of
  Vector Dark Matter}}, \href{https://doi.org/10.1143/PTP.126.435}{\emph{Prog.
  Theor. Phys.} {\bfseries 126} (2011) 435}
  [\href{https://arxiv.org/abs/1012.5455}{{\ttfamily 1012.5455}}].

\bibitem{Davoudiasl:2013jma}
H.~Davoudiasl and I.~M. Lewis, \emph{{Dark Matter from Hidden Forces}},
  \href{https://doi.org/10.1103/PhysRevD.89.055026}{\emph{Phys. Rev. D}
  {\bfseries 89} (2014) 055026}
  [\href{https://arxiv.org/abs/1309.6640}{{\ttfamily 1309.6640}}].

\bibitem{Gross:2015cwa}
C.~Gross, O.~Lebedev and Y.~Mambrini, \emph{{Non-Abelian gauge fields as dark
  matter}}, \href{https://doi.org/10.1007/JHEP08(2015)158}{\emph{JHEP}
  {\bfseries 08} (2015) 158}
  [\href{https://arxiv.org/abs/1505.07480}{{\ttfamily 1505.07480}}].

\bibitem{Karam:2015jta}
A.~Karam and K.~Tamvakis, \emph{{Dark matter and neutrino masses from a
  scale-invariant multi-Higgs portal}},
  \href{https://doi.org/10.1103/PhysRevD.92.075010}{\emph{Phys. Rev. D}
  {\bfseries 92} (2015) 075010}
  [\href{https://arxiv.org/abs/1508.03031}{{\ttfamily 1508.03031}}].

\bibitem{Choi:2019zeb}
S.-M. Choi, H.~M. Lee, Y.~Mambrini and M.~Pierre, \emph{{Vector SIMP dark
  matter with approximate custodial symmetry}},
  \href{https://doi.org/10.1007/JHEP07(2019)049}{\emph{JHEP} {\bfseries 07}
  (2019) 049} [\href{https://arxiv.org/abs/1904.04109}{{\ttfamily
  1904.04109}}].

\bibitem{Elahi:2019jeo}
F.~Elahi and S.~Khatibi, \emph{{Multi-Component Dark Matter in a Non-Abelian
  Dark Sector}}, \href{https://doi.org/10.1103/PhysRevD.100.015019}{\emph{Phys.
  Rev. D} {\bfseries 100} (2019) 015019}
  [\href{https://arxiv.org/abs/1902.04384}{{\ttfamily 1902.04384}}].

\bibitem{Abe:2020mph}
T.~Abe, M.~Fujiwara, J.~Hisano and K.~Matsushita, \emph{{A model of
  electroweakly interacting non-abelian vector dark matter}},
  \href{https://arxiv.org/abs/2004.00884}{{\ttfamily 2004.00884}}.

\bibitem{Nugaev:2020zcv}
E.~Nugaev and A.~Shkerin, \emph{{Unveiling complex vector dark matter by
  magnetic field}},  \href{https://arxiv.org/abs/2004.14354}{{\ttfamily
  2004.14354}}.

\bibitem{Elahi:2020urr}
F.~Elahi and M.~Mohammadi~Najafabadi, \emph{{Neutron Decay to a Non-Abelian
  Dark Sector}},  \href{https://arxiv.org/abs/2005.00714}{{\ttfamily
  2005.00714}}.

\bibitem{Hagiwara:1986vm}
K.~Hagiwara, R.~D. Peccei, D.~Zeppenfeld and K.~Hikasa, \emph{{Probing the Weak
  Boson Sector in $e^+e^-\to W^+W^-$}},
  \href{https://doi.org/10.1016/0550-3213(87)90685-7}{\emph{Nucl. Phys.}
  {\bfseries B282} (1987) 253}.

\bibitem{Gaemers:1978hg}
K.~Gaemers and G.~Gounaris, \emph{{Polarization Amplitudes for $e^+e^- \to
  W^+W^-$ and $e^+e^- \to ZZ$}},
  \href{https://doi.org/10.1007/BF01440226}{\emph{Z. Phys. C} {\bfseries 1}
  (1979) 259}.

\bibitem{Gounaris:1996rz}
G.~Gounaris et~al., \emph{{Triple gauge boson couplings}},  in \emph{{AGS /
  RHIC Users Annual Meeting}}, pp.~525--576, 1, 1996,
  \href{https://arxiv.org/abs/hep-ph/9601233}{{\ttfamily hep-ph/9601233}}.

\bibitem{Helm:1956zz}
R.~H. Helm, \emph{{Inelastic and Elastic Scattering of 187-Mev Electrons from
  Selected Even-Even Nuclei}},
  \href{https://doi.org/10.1103/PhysRev.104.1466}{\emph{Phys. Rev.} {\bfseries
  104} (1956) 1466}.

\bibitem{Lewin:1995rx}
J.~D. Lewin and P.~F. Smith, \emph{{Review of mathematics, numerical factors,
  and corrections for dark matter experiments based on elastic nuclear
  recoil}},
  \href{https://doi.org/10.1016/S0927-6505(96)00047-3}{\emph{Astropart. Phys.}
  {\bfseries 6} (1996) 87}.

\bibitem{Chang:2010en}
S.~Chang, N.~Weiner and I.~Yavin, \emph{{Magnetic Inelastic Dark Matter}},
  \href{https://doi.org/10.1103/PhysRevD.82.125011}{\emph{Phys. Rev.}
  {\bfseries D82} (2010) 125011}
  [\href{https://arxiv.org/abs/1007.4200}{{\ttfamily 1007.4200}}].

\bibitem{Aprile:2018dbl}
{\scshape XENON} collaboration, \emph{{Dark Matter Search Results from a One
  Ton-Year Exposure of XENON1T}},
  \href{https://doi.org/10.1103/PhysRevLett.121.111302}{\emph{Phys. Rev. Lett.}
  {\bfseries 121} (2018) 111302}
  [\href{https://arxiv.org/abs/1805.12562}{{\ttfamily 1805.12562}}].

\bibitem{Agnese:2014aze}
{\scshape SuperCDMS} collaboration, \emph{{Search for Low-Mass Weakly
  Interacting Massive Particles with SuperCDMS}},
  \href{https://doi.org/10.1103/PhysRevLett.112.241302}{\emph{Phys. Rev. Lett.}
  {\bfseries 112} (2014) 241302}
  [\href{https://arxiv.org/abs/1402.7137}{{\ttfamily 1402.7137}}].

\bibitem{Abdelhameed:2019hmk}
{\scshape CRESST} collaboration, \emph{{First results from the CRESST-III
  low-mass dark matter program}},
  \href{https://doi.org/10.1103/PhysRevD.100.102002}{\emph{Phys. Rev. D}
  {\bfseries 100} (2019) 102002}
  [\href{https://arxiv.org/abs/1904.00498}{{\ttfamily 1904.00498}}].

\bibitem{Amole:2019fdf}
{\scshape PICO} collaboration, \emph{{Dark Matter Search Results from the
  Complete Exposure of the PICO-60 C$_3$F$_8$ Bubble Chamber}},
  \href{https://doi.org/10.1103/PhysRevD.100.022001}{\emph{Phys. Rev.}
  {\bfseries D100} (2019) 022001}
  [\href{https://arxiv.org/abs/1902.04031}{{\ttfamily 1902.04031}}].

\bibitem{Aprile:2015uzo}
{\scshape XENON} collaboration, \emph{{Physics reach of the XENON1T dark matter
  experiment}},
  \href{https://doi.org/10.1088/1475-7516/2016/04/027}{\emph{JCAP} {\bfseries
  1604} (2016) 027} [\href{https://arxiv.org/abs/1512.07501}{{\ttfamily
  1512.07501}}].

\bibitem{inprep}
In preparation.

\bibitem{Lahanas:1994dv}
A.~B. Lahanas and V.~C. Spanos, \emph{{Static quantities of the W boson in the
  MSSM}}, \href{https://doi.org/10.1016/0370-2693(94)90703-X}{\emph{Phys.
  Lett.} {\bfseries B334} (1994) 378}
  [\href{https://arxiv.org/abs/hep-ph/9405298}{{\ttfamily hep-ph/9405298}}].

\bibitem{Aprile:2011hx}
{\scshape XENON100} collaboration, \emph{{Likelihood Approach to the First Dark
  Matter Results from XENON100}},
  \href{https://doi.org/10.1103/PhysRevD.84.052003}{\emph{Phys. Rev.}
  {\bfseries D84} (2011) 052003}
  [\href{https://arxiv.org/abs/1103.0303}{{\ttfamily 1103.0303}}].

\bibitem{Athron:2018hpc}
{\scshape GAMBIT} collaboration, \emph{{Global analyses of Higgs portal singlet
  dark matter models using GAMBIT}},
  \href{https://doi.org/10.1140/epjc/s10052-018-6513-6}{\emph{Eur. Phys. J.}
  {\bfseries C79} (2019) 38}
  [\href{https://arxiv.org/abs/1808.10465}{{\ttfamily 1808.10465}}].

\bibitem{Athron:2018ipf}
P.~Athron, J.~M. Cornell, F.~Kahlhoefer, J.~Mckay, P.~Scott and S.~Wild,
  \emph{{Impact of vacuum stability, perturbativity and XENON1T on global fits
  of $\mathbb {Z}_2$ and $\mathbb {Z}_3$ scalar singlet dark matter}},
  \href{https://doi.org/10.1140/epjc/s10052-018-6314-y}{\emph{Eur. Phys. J.}
  {\bfseries C78} (2018) 830}
  [\href{https://arxiv.org/abs/1806.11281}{{\ttfamily 1806.11281}}].

\bibitem{Aprile:2015lha}
{\scshape XENON} collaboration, \emph{{Lowering the radioactivity of the
  photomultiplier tubes for the XENON1T dark matter experiment}},
  \href{https://doi.org/10.1140/epjc/s10052-015-3657-5}{\emph{Eur. Phys. J.}
  {\bfseries C75} (2015) 546}
  [\href{https://arxiv.org/abs/1503.07698}{{\ttfamily 1503.07698}}].

\bibitem{Barrow:2016doe}
P.~Barrow et~al., \emph{{Qualification Tests of the R11410-21 Photomultiplier
  Tubes for the XENON1T Detector}},
  \href{https://doi.org/10.1088/1748-0221/12/01/P01024}{\emph{JINST} {\bfseries
  12} (2017) P01024} [\href{https://arxiv.org/abs/1609.01654}{{\ttfamily
  1609.01654}}].

\bibitem{DelNobile:2013sia}
M.~Cirelli, E.~Del~Nobile and P.~Panci, \emph{{Tools for model-independent
  bounds in direct dark matter searches}},
  \href{https://doi.org/10.1088/1475-7516/2013/10/019}{\emph{JCAP} {\bfseries
  1310} (2013) 019} [\href{https://arxiv.org/abs/1307.5955}{{\ttfamily
  1307.5955}}].

\bibitem{Witte:2017qsy}
S.~J. Witte and G.~B. Gelmini, \emph{{Updated Constraints on the Dark Matter
  Interpretation of CDMS-II-Si Data}},
  \href{https://doi.org/10.1088/1475-7516/2017/05/026}{\emph{JCAP} {\bfseries
  1705} (2017) 026} [\href{https://arxiv.org/abs/1703.06892}{{\ttfamily
  1703.06892}}].

\bibitem{Yellin:2002xd}
S.~Yellin, \emph{{Finding an upper limit in the presence of unknown
  background}}, \href{https://doi.org/10.1103/PhysRevD.66.032005}{\emph{Phys.
  Rev.} {\bfseries D66} (2002) 032005}
  [\href{https://arxiv.org/abs/physics/0203002}{{\ttfamily physics/0203002}}].

\bibitem{Ferrer:2015bta}
F.~Ferrer, A.~Ibarra and S.~Wild, \emph{{A novel approach to derive
  halo-independent limits on dark matter properties}},
  \href{https://doi.org/10.1088/1475-7516/2015/09/052}{\emph{JCAP} {\bfseries
  1509} (2015) 052} [\href{https://arxiv.org/abs/1506.03386}{{\ttfamily
  1506.03386}}].

\bibitem{DDCalc:url}
\url{http://ddcalc.hepforge.org/}.

\bibitem{DarkBit:2017lvb}
{\scshape The GAMBIT Dark Matter Workgroup} collaboration, \emph{{DarkBit: A
  GAMBIT module for computing dark matter observables and likelihoods}},
  \href{https://doi.org/10.1140/epjc/s10052-017-5155-4}{\emph{Eur. Phys. J.}
  {\bfseries C77} (2017) 831}
  [\href{https://arxiv.org/abs/1705.07920}{{\ttfamily 1705.07920}}].

\bibitem{Mancuso:2018zoh}
M.~Mancuso et~al., \emph{{A Low Nuclear Recoil Energy Threshold for Dark Matter
  Search with CRESST-III Detectors}},
  \href{https://doi.org/10.1007/s10909-018-1948-6}{\emph{J. Low. Temp. Phys.}
  {\bfseries 193} (2018) 441}.

\bibitem{Abdelhameed:2019mac}
{\scshape CRESST} collaboration, \emph{{Description of CRESST-III Data}},
  \href{https://arxiv.org/abs/1905.07335}{{\ttfamily 1905.07335}}.

\bibitem{Patel:2015tea}
H.~H. Patel, \emph{{Package-X: A Mathematica package for the analytic
  calculation of one-loop integrals}},
  \href{https://doi.org/10.1016/j.cpc.2015.08.017}{\emph{Comput. Phys. Commun.}
  {\bfseries 197} (2015) 276}
  [\href{https://arxiv.org/abs/1503.01469}{{\ttfamily 1503.01469}}].

\end{thebibliography}\endgroup
\end{document}